\documentclass[apj]{emulateapj}
\usepackage{apjfonts}
\usepackage{amsbsy}

\usepackage{color}
\usepackage{rotating}

\def\hmpc{$h^{-1}$Mpc}

\def\msol{M$_\odot$}

\def\rvir{R_{\rm vir}}

\def\nsat{\langle N_{\rm sat}\rangle_M}
\def\ncen{\langle N_{\rm cen}\rangle_M}

\def\mcut{M_{\rm cut}}
\def\msat{M_{\rm sat}}

\def\asat{\alpha_{\rm sat}}

\def\om{\Omega_m}

\def\omb{\Omega_b}
\def\s8{\sigma_8}

\def\lcdm{$\Lambda$CDM}

\def\rhocrit{\rho_{\rm crit}}

\def\x2{$\chi^2$}

\def\NNm1{\langle N(N-1) \rangle}

\def\fsat{f_{\rm sat}}

\def\m_star{M_\ast}

\def\lcdm{$\Lambda$CDM}

\def\om{\Omega_m}

\def\omb{\Omega_b}
\def\s8{\sigma_8}

\def\hmpc{$h^{-1}\,$Mpc}

\def\x2{$\chi^2$}

\def\mcut{M_{\rm cut}}

\def\nsat{\langle N_{\mbox{\scriptsize sat}}\rangle_M}

\def\ncen{\langle N_{\mbox{\scriptsize cen}}\rangle_M}

\def\NNm1{\langle N(N-1) \rangle}

\def\fsat{f_{\rm sat}}

\def\fsat{f_{\rm sat}}

\def\p0{P_0(r)}

\def\msat{M_{\rm sat}}

\def\fred{f_{\rm red}}

\def\msol{$M$_\odot}
\def\m12{M_{12}}

\def\m12{M_{12}}

\def\rs0{\hat{R}_{\rm sh}^0}
\def\mg2{Mg\,II}

\def\fsat{f_{\rm sat}}

\def\fq{f_{\rm Q}}
\def\tq{t_{\rm Q}}

\def\mgal{M_{\ast}}

\def\DS{\Delta\Sigma}

\def\mhalo{M_h}
\def\sigsm{\sigma_{\rm log M_{\ast}}}
\def\fshmr{f_{\rm SHMR}}
\def\fshmri{f_{\rm \small SHMR}^{-1}}
\def\Bsat{B_{\rm sat}}
\def\bsat{\beta_{\rm sat}}
\def\Bcut{B_{\rm cut}}
\def\bcut{\beta_{\rm cut}}
\def\wtheta{w_{\theta}}

\def\fred{f_{q}}

\begin{document}

\title{Evolution of the Stellar-to-Dark Matter Relation: \\Separating
  Star-Forming and Passive Galaxies from $z=1$ to 0}

\author{Jeremy L. Tinker\altaffilmark{1}, 
Alexie Leauthaud\altaffilmark{2}, 
Kevin Bundy\altaffilmark{2},
Matthew  R. George\altaffilmark{3}, \\
Peter Behroozi\altaffilmark{4},
Richard Massey\altaffilmark{5},
Jason Rhodes\altaffilmark{6,7},
Risa H. Wechsler\altaffilmark{4}
}

%\submitted{Submitted to ApJL}
\email{jeremy.tinker@nyu.edu}

\altaffiltext{1}{Center for Cosmology and Particle Physics, Department
  of Physics, New York University}

\altaffiltext{2}{Kavli Institute for the Physics and Mathematics of
  the Universe, Todai Institutes for Advanced Study, the University of Tokyo, Kashiwa, Japan 277-8583 (Kavli IPMU, WPI)}

\altaffiltext{3}{Department of Astronomy, University of California,
  and Lawrence Berkeley National Laboratory,
  Berkeley, CA 94720, USA}
\altaffiltext{4}{Kavli Institute for Particle Astrophysics and
 Cosmology; Physics Department, Stanford University, and SLAC
 National Accelerator Laboratory, Stanford CA 94305}

\altaffiltext{5}{Institute for Computational Cosmology, Durham University, South Road, Durham, DH1 3LE, U.K.}

\altaffiltext{6}{California Institute of Technology, MC 350-17, 1200
 East California Boulevard, Pasadena, CA 91125, USA}

\altaffiltext{7}{Jet Propulsion Laboratory, California Institute of Technology, Pasadena, CA 91109}

\begin{abstract}

  We use measurements of the stellar mass function, galaxy clustering,
  and galaxy-galaxy lensing within the COSMOS survey to constrain the
  stellar-to-halo mass relation (SHMR) of star forming and quiescent
  galaxies over the redshift range $z=[0.2,1.0]$. For massive
  galaxies, $\mgal\gtrsim 10^{10.6} \msol$, our results indicate that
  star-forming galaxies grow proportionately as fast as their dark
  matter halos while quiescent galaxies are outpaced by dark matter
  growth. At lower masses, there is minimal difference in the SHMRs,
  implying that the majority low-mass quiescent galaxies have only
  recently been quenched of their star formation. Our analysis also
  affords a breakdown of all COSMOS galaxies into the relative numbers
  of central and satellite galaxies for both populations.  At $z=1$,
  satellite galaxies dominate the red sequence below the knee in the
  stellar mass function. But the number of quiescent satellites
  exhibits minimal redshift evolution; all evolution in the red
  sequence is due to low-mass central galaxies being quenched of their
  star formation. At $\mgal\sim 10^{10} \msol$, the fraction of
  central galaxies on the red sequence increases by a factor of ten
  over our redshift baseline, while the fraction of quenched satellite
  galaxies at that mass is constant with redshift. We define a
  ``migration rate'' to the red sequence as the time derivative of the
  passive galaxy abundances.  We find that the migration rate of
  central galaxies to the red sequence increases by nearly an order of
  magnitude from $z=1$ to $z=0$. These results imply that the
  efficiency of quenching star formation for centrals is increasing
  with cosmic time, while the mechanisms that quench the star
  formation of satellite galaxies in groups and clusters is losing
  efficiency.

\end{abstract}


\section{Introduction}

One of the defining characteristics of the $z=0$ galaxy distribution
is its bimodality. Galaxies can be roughly categorized into the
star-forming sequence of blue, disky, gas-rich galaxies, and the
quiescent, ellipsoidal galaxies with old stellar populations and red
colors (\citealt{strateva_etal:01, blanton_etal:03cmd,
  kauffmann_etal:03b, madgwick_etal:03}). This bimodality is firmly in
place at $z=1$ (\citealt{bell_etal:04, cooper_etal:06,
  willmer_etal:06}) and extends out to $z=2$ and possibly beyond
(\citealt{kriek_etal:08, williams_etal:09}). The physical processes
that drive the creation and evolution of the red sequence are not
fully understood. There are many possible routes to the red sequence,
but the relative efficiency of each are unquantified. In this paper we
use measurements of the stellar mass function, galaxy clustering, and
galaxy-galaxy lensing from the COSMOS survey
(\citealt{scoville_etal:07}) to disentangle the various process that
attenuate star formation in galaxies. This paper is an extension of
\cite{leauthaud_etal:11a, leauthaud_etal:12_shmr} (hereafter, L11 and
L12). In L11 we presented our theoretical framework; in L12 we applied
this framework to stellar mass defined samples in COSMOS; in this
paper we extend this framework to samples defined by both stellar mass
and star formation activity, and apply it once again to COSMOS data.

\begin{figure*}%{figure}
%\epsscale{1.0} 
\plotone{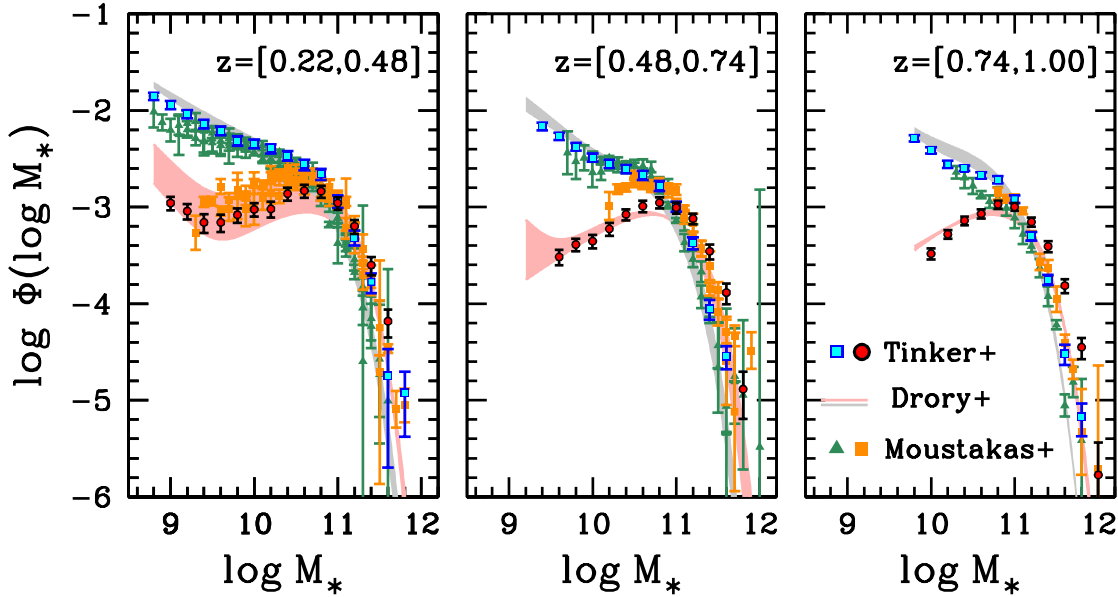}
\vspace{-6.cm}
\caption{ \label{smf_compare} The COSMOS stellar mass function,
  measured in our redshift bins, compared to other stellar mass
  functions at comparable redshifts. The blue squares and red circles
  represent our COSMOS measurements for SF and passive galaxies,
  respectively. The gray and pink shaded bands show the COSMOS
  measurements from \cite{drory_etal:09}. Because Drory measured the
  SMF in different redshift bins, the bands show the range of SMF
  values for the two bins that overlap with each redshift bin used
  here. The orange squares and green triangles represent the
  measurements from PRIMUS (\citealt{moustakas_etal:13}) for passive
  and SF galaxies. In each panel, PRIMUS results are shown for all
  redshift bins whose median redshift is contained within the given
  COSMOS redshift bin (shown at the top of each panel). PRIMUS results
  are based on spectroscopy, thus do not go as faint as the two COSMOS
  results.}
\end{figure*}%{figure}

\begin{figure}
\epsscale{1.2}
\plotone{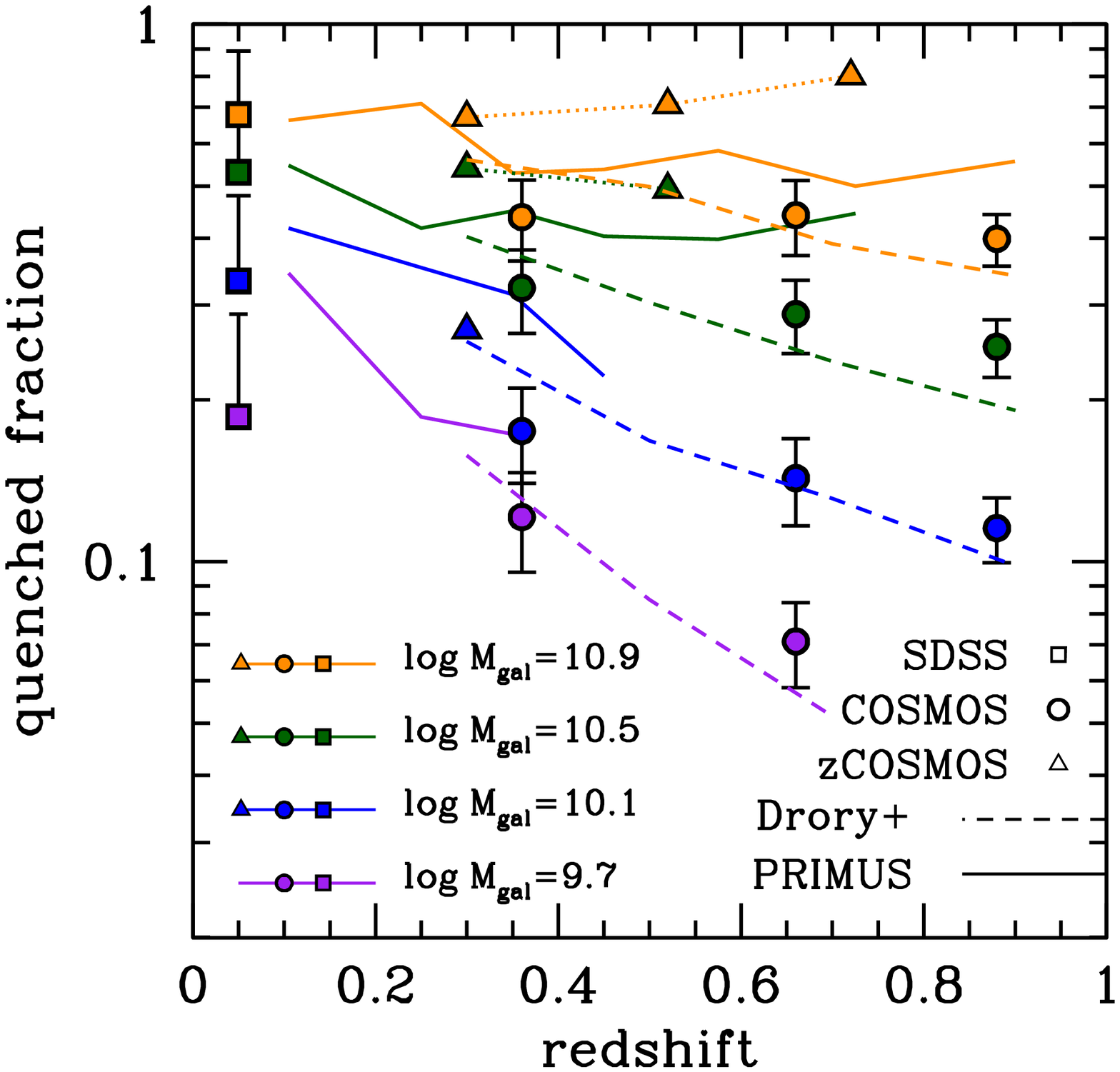}
\caption{\label{cosmos_primus_redfrac} A comparison of the evolving
  quenched fractions in COSMOS (from this paper), COSMOS (from
  \citealt{drory_etal:09}), PRIMUS (\citealt{moustakas_etal:13}), and
  zCOSMOS (\citealt{knobel_etal:13}) and SDSS. Circles with errors are the
  COSMOS data in this paper from Figure \ref{bestfit_smf}, which use a
  $NUV-R-J$ color diagram to isolate passive galaxies. Squares are
  from a volume-limited SDSS sample that uses $D_n4000$ to determine
  $\fred$. This sample will be used later in the paper and is
  discussed in \S 4.3. The dashed lines show the results from the
  \cite{drory_etal:09} COSMOS measurements, which are from the same
  raw data but use different methods to determine stellar mass. The
  PRIMUS results use a different estimate of stellar mass and use SED
  fitting to determine the delineation between passive and active
  galaxies. The zCOSMOS results use a single $U-B$ color cut to
  determine the set of passive galaxies. }
\end{figure}

\begin{figure*}%{figure}
%\epsscale{1.0} 
\plotone{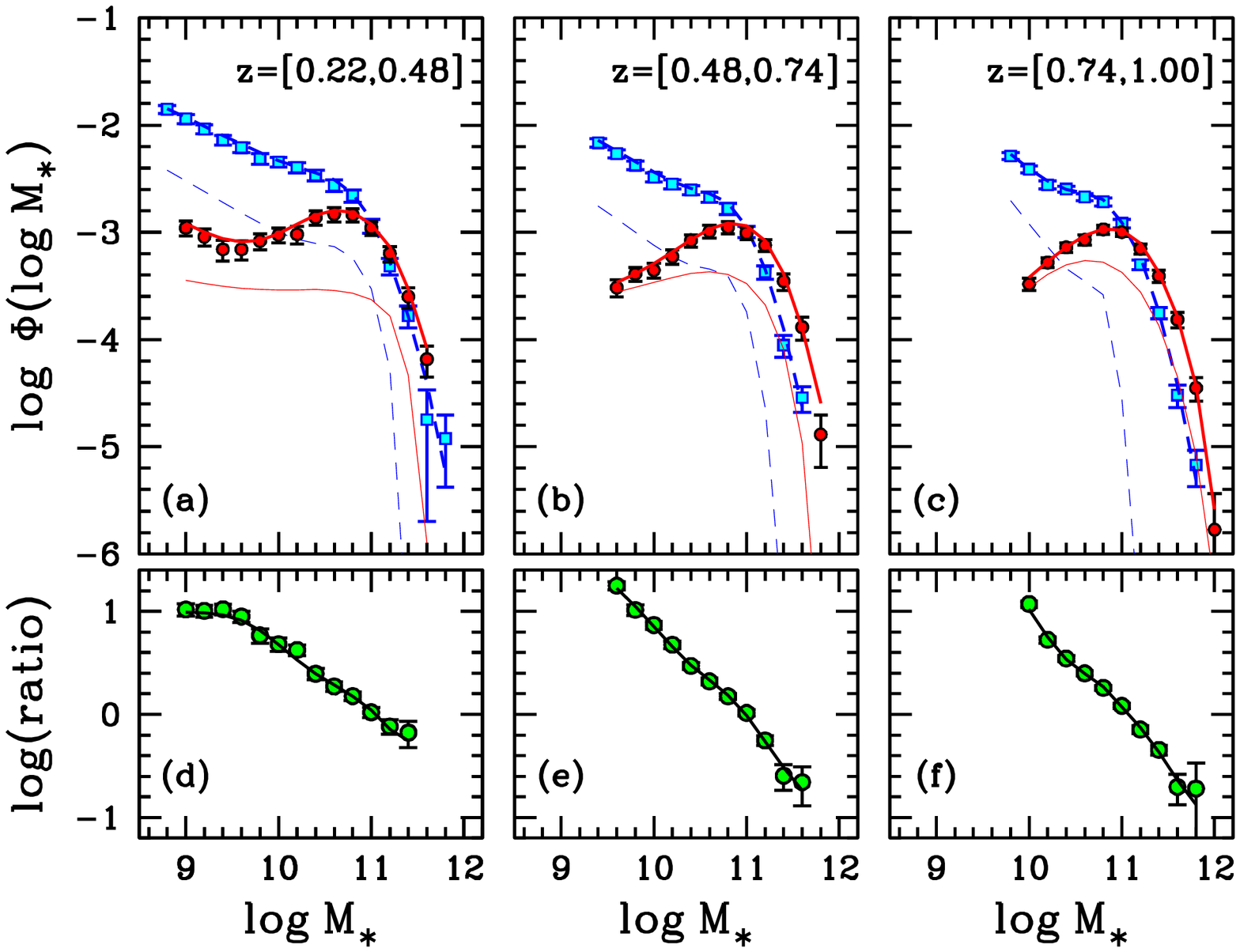}
\vspace{-3.cm}
\caption{ \label{bestfit_smf} {\it Upper panels:} The stellar mass
  functions in each redshift bin broken down by star formation
  activity. Points with errors represent COSMOS measurements; curves
  represent best-fit HOD models. Red circles represent passive galaxies
  while blue squares represent SF galaxies. Error bars are obtained
  from mock galaxy samples discussed in \S \ref{s.data}. Red solid
  curves represent the HOD model for passive galaxies. The dashed blue
  curves represent the HOD model for SF galaxies. The thin dashed
  and solid curves in each panel represent the abundance of satellite
  galaxies only for each subsample. {\it Lower panels:} The
  red-to-blue ratio of the SMFs. This quantity contains complementary
  information to the individual SMFs because the amplitudes of the passive
  and SF SMFs are correlated. Points with errors represent the
  COSMOS measurements, while the black curve is the HOD model. }
\end{figure*}%{figure}

The proposed mechanisms for quenching star formation can be grouped
into two broad categories: processes that affect galaxies that exist
at the center of the potential well of their host dark matter halo,
and processes that affect galaxies that orbit as satellites within a
larger dark matter potential. Central galaxy processes include
mergers, AGN feedback--- triggered either by mergers or by disk
instabilities---and shock heating of infalling gas at a critical halo
mass scale or galaxy mass scale (e.g., \citealt{croton_etal:06a,
  bower_etal:06, dekel_birnboim:06, cattaneo_etal:06,
  hopkins_etal:08b}). Satellite galaxy processes do include some AGN
and merging activity, but are likely dominated by tidal effects from
the host dark matter halo, harassment by other galaxies within the
group, strangulation from an active cold gas supply, and ram pressure
stripping of gas by interaction with the host halo's hot gas (e.g.,
\citealt{gunn_gott:72, moore_etal:98, balogh_etal:00}).

In this work we define a galaxy group as a set of galaxies that share
a common dark matter halo. Close pairs of halos certainly exist in the
field (e.g., the Milky Way-Andromeda pair), but by our definition
these are not galaxy groups. This definition matches up to the
division of processes that quench galaxies defined in the previous
paragraph: ram pressure, tidal stripping, and strangulation do not
significantly affect galaxies until they have crossed the virial
radius of a larger halo\footnote{Some studies have found an increased
  fraction of quenched galaxies extending several virial radii outside
  of clusters (e.g., \citealt{balogh_etal:00, hansen_etal:09,
    vonderlinden_etal:10}), but these results are easily explained by
  accounting for galaxies that are satellites in nearby groups, as
  well as galaxies in the cluster infall region that have orbited
  within the virial radius of the cluster, but the apocenter of their
  orbit is outside $\rvir$ (\citealt{wetzel_etal:13_groups3}.)}. This
definition also fits seamlessly with our theoretical framework for
analyzing the clustering and lensing of galaxies.

To disentangle the relative numbers of central and satellite galaxies,
we use the framework of the Halo Occupation Distribution (HOD; see,
e.g., \citealt{peacock_smith:00, seljak:00, roman_etal:01,
  cooray_sheth:02, berlind_weinberg:02} for early works, and
\citealt{zheng_etal:07, vdb_etal:07, tinker_etal:10_drg} for examples
of more recent implementations of the framework). In brief, the HOD
provides a statistical framework for the probability distribution
function of galaxies within halos. Traditionally, HOD models
parameterize $P(N|\mhalo)$, the probability that a halo of mass $M$
contains $N$ galaxies in a pre-defined sample. The HOD for a given
galaxy mass, $\mgal$, is based on two characteristic halo mass scales:
the mean halo mass for central galaxies and a larger halo mass where
there is (on average) one satellite galaxy of mass $M\ge\mgal$. Here we
use an extended model that parameterizes this probability as a
function of galaxy mass, $P(N|\mhalo,\mgal)$, rather than for a
specified threshold. The specific model we implement is described in
detail in L11, which begins with parameterization of the stellar
mass-to-halo mass relation for central galaxies (SHMR). This function
specifies the mean mass of a central galaxy as a function of halo
mass. The halo mass scale for satellite galaxies is motivated by
previous HOD analyses that find a tight relation between these two
halo mass scales.

The benefit of the COSMOS survey for this work is that it provides a
consistent set of observations and the same definition of stellar mass
at various redshifts. Additionally, the broad wavelength coverage of
COSMOS are highly efficient at differentiating dusty star-forming
galaxies from truly passive objects. Although data exists at multiple
epochs from various surveys, the clustering of galaxies depend
sensitively on survey selection (\citealt{sanchez_cole:08}), and
stellar mass estimates depend on both survey parameters and on
assumptions in the stellar mass modeling (\citealt{conroy_etal:09,
  conroy_gunn:10}). Constraints on the redshift evolution in the SHMR
are significantly weakened when incorporating such uncertainties into
the analysis (\citealt{behroozi_etal:10}), thus the COSMOS data set is
crucial for identifying true redshift trends.

In this paper we will interchangeably use the terms ``quiescent'',
``quenched'', and ``passive'' to refer to galaxies that have little to
no star formation and are intrinsically located on the red
sequence. Galaxies that appear red due to dust contamination of
broadband colors are included in the star-forming sequence.  We will
discuss this further in \S 2. We will refer to the ``red sequence'' to
mean the set of galaxies that are intrinsically red. The complement of
the red sequence is the set of star-forming (SF) galaxies.We will use
the knee in the stellar mass function, approximately
$10^{10.6}\,\msol$ at all redshifts considered
(\citealt{drory_etal:09, marchesini_etal:09}), as the reference point
between ``high-mass'' and ``low-mass'' galaxy samples. Our reference
point for small and large distance scales is 1 Mpc (comoving; $\sim
110$ arcsec at $z=0.5$), which is the at the center of the transition
in the galaxy correlation function from pair counts being dominated by
galaxies in two distinct halos and pairs that arise from two galaxies
occupying a common halo. We will frequently refer to both the fraction
of galaxies that are satellites, $\fsat$, the fraction of galaxies
that are quenched $\fred$, and combinations of both. For clarity, the
fraction of satellites that are quenched is referenced as $\fred(sat)$
while the fraction of quenched galaxies that are satellites is
referenced as $\fsat(q)$---i.e., the subsample {\it for which} the
fraction is determined is referenced parenthetically, while the
quantity {\it by which} the fraction is determined is listed in the
subscript.

In all theoretical modeling we assume a flat \lcdm\ cosmological model
of ($\om$,$\s8$,$\omb$,$n_s$,$h_0) =
(0.272,0.807,0.0438,0.963,0.72)$. We define a dark matter halo as a
spherical, virialized object with a mean interior density of
$\Delta\equiv 3\mhalo/4\pi\om\rhocrit R_{h}^3=200$. All halo
statistics used in this paper are calibrated from numerical
simulations that match this halo definition.

\begin{figure*}%{figure}
%\epsscale{1.0} 
\plotone{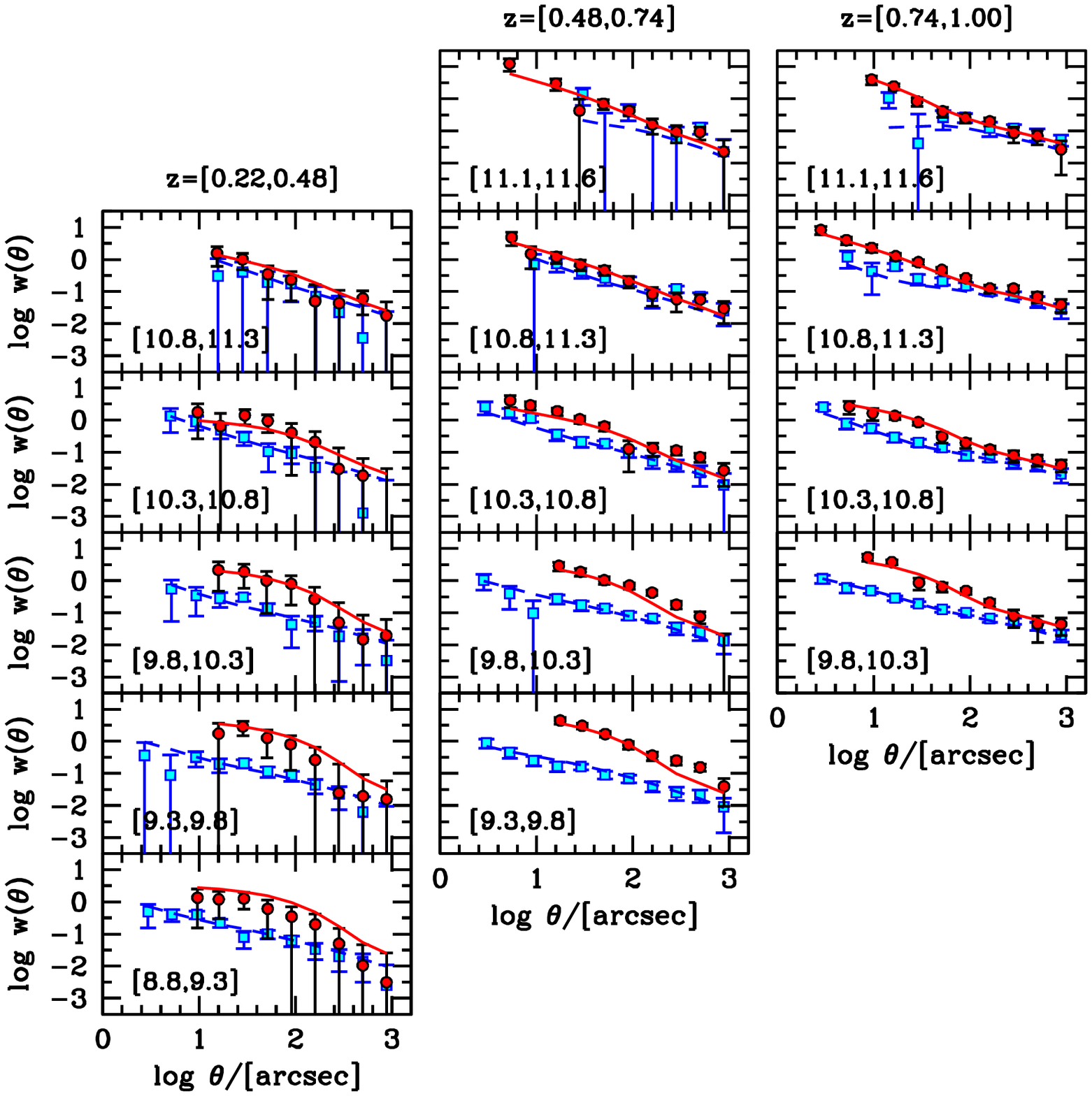}
%\vspace{-7.5cm}
\caption{ \label{bestfit_wtheta} Angular clustering of COSMOS galaxies
  in stellar mass bins. From left to right, columns represent
  measurements at $z=0.36$, $z=0.66$, and $z=0.88$. Point with error
  bars are measurements while curves indicate best-fit HOD
  models. Colors and point types are the same as Figure
  \ref{bestfit_smf}. Only angular bins with more than 10 pairs
  used in the analysis, thus data for passive galaxies often do not extend
  to the minimum angular separation. The volume of each redshift bin
  depends strongly on the median redshift, as indicated in Table
  1. Thus, the $z=0.36$ measurements have the largest error bars
  because they are taken from the smallest volume. For mass bins at
  $\log\mgal\le 10.3$, the enhanced clustering of passive galaxies is
  driven by the high fraction of satellite galaxies that are quenched
  (c.f., Figure \ref{fred_evolve}.)}
\end{figure*}%{figure}

% **************** TABLE 1 ***************
% Number of Galaxies
% \input{tab1.tex}

% **************** TABLE 1 ***************
% GG-lensing bins
\begin{deluxetable}{lccccccc}
%  \tabletypesize{\scriptsize} 
\tablecolumns{8} \tablecaption{Binning scheme for galaxies in $\log_{10}(\mgal)$ \label{gg_binning}} \tablewidth{0pt} 
\startdata
\hline 
\hline 
\\  [-1.5ex]
 $\DS$  &  bin1 &  bin2 &  bin3 &  bin4 &  bin5 &  bin6 &  bin7 \\ [1ex]
\hline\\  [-1.5ex]
$z_1$, min & 11.12 & 10.89 & 10.64 & 10.3 & 9.82 & 9.2 & 8.7 \\
$z_1$, max & 12.0 & 11.12 & 10.89 & 10.64 & 10.3 & 9.8 & 9.2 \\
\hline
$z_2$, min & 11.29 & 11.05 & 10.88 & 10.65 & 10.3 & 9.8 & 9.3 \\
$z_2$, max & 12.0 & 11.29 & 11.05 & 10.88 & 10.65 & 10.3 & 9.8 \\
\hline
$z_3$, min & 11.35 & 11.16 & 10.97 & 10.74 & 10.39 & 9.8 & --- \\
$z_3$, max & 12.0 & 11.35 & 11.16 & 10.97 & 10.74 & 10.39 & --- 
\\  
\\
$\wtheta$   &  bin1 &  bin2 &  bin3 &  bin4
   &  bin5 &  bin6 & \\ [1ex]
\hline\\  [-1.5ex]

$z_1$, min & 8.8 & 9.3 & 9.8 & 10.3 & 10.8 &---  & \\
$z_1$, max & 9.3 & 9.8 & 10.3 & 10.8 & 11.3 & --- & \\
\hline
$z_2$, min & --- & 9.3 & 9.8 & 10.3 & 10.8 & 11.1  & \\
$z_2$, max & --- & 9.8 & 10.3 & 10.8 & 11.3 & 11.6 & \\
\hline
$z_3$, min & --- & --- & 9.8 & 10.3 & 10.8 & 11.1  & \\
$z_3$, max & --- & --- & 10.3 & 10.8 & 11.3 & 11.6 & \\
 [-1.5ex]

\enddata
\end{deluxetable}

\section{Data}\label{s.data}

Details of the COSMOS survey can be found in
\cite{scoville_etal:07}. Details of the measurement techniques and
methods for the stellar mass functions (SMFs), angular galaxy
clustering ($\wtheta$), and galaxy-galaxy lensing ($\DS$) can be found
in L12. All $\wtheta$ measurements are taken from the Subaru catalog
(2.3 deg$^2$) while lensing and SMF measurements are restricted to the
$HST$ ACS catalog (1.64 deg$^2$). The sample selection is also
identical to L12. Here we repeat all these measurements, now broken
into two subsamples of star-forming (blue) and passive (red)
objects. Intrinsically passive galaxies are identified in a specific
region in the $(NUV-R)-(R-J)$ color-color space in the same manner as
\cite{bundy_etal:10}). The addition of near-IR data breaks the
degeneracy between dusty and star-forming objects
(\citealt{pozzetti_mannucci:00, labbe_etal:05, williams_etal:09,
  zhu_etal:11}).

Photometric redshifts are obtained from \cite{ilbert_etal:09},
versions v1.7 and v1.8. These photo-$z$ estimates have negligible
differences at $z<1$ but v1.8 has improved accuracy relative to
spectroscopic redshifts. The v1.7 photo-$z$'s are used for the SMF and
$\wtheta$ and v1.8 for the lensing catalog. Later version of the
photometric redshifts, which were not available during much of the
present work, focus on $z\sim 2$. We have confirmed that there are
negligible changes to $z<1$ results.

Stellar masses are estimated using the Bayesian code of
\cite{bundy_etal:06} and assume a \cite{chabrier:03} IMF. While the method of Bundy et.~al.~uses multi-band colors to constrain the $M/L$ ratio in the observed K-band, with mass estimates derived from application to the K-band luminosity only.

We restrict
our analysis to stellar masses above the 80\% stellar mass
completeness limit, as in L12. Due to the lower intrinsic luminosities
at fixed stellar mass, passive galaxies have a higher completeness
limit at fixed redshift by roughly $\sim 0.2$ dex. For the stellar
mass function and the galaxy-galaxy lensing measurements, we restrict
the measurements of the passive population to be above that limit
(although we will compare our lensing fits to the data for these bins
in the presentation of our results for comparison purposes). For the
clustering measurements, we find that including passive galaxies down
to the star-forming stellar mass limit does not bias the clustering of
those samples\footnote{The passive galaxy completeness limit
  essentially cuts part-way through the lowest stellar mass bin at
  each redshift interval. We compares measurements of the clustering
  of all passive galaxies in this bin to those that are above the
  limit, finding that the results are consistent with one another, but
  the higher number of galaxies in the full bin yields better error
  bars, especially at small scales.}  thus we incorporate these
galaxies in the clustering bins. See Figure 2 in L12 for a plot of the
completeness limits as a function of redshift. Stellar mass limits are
also given in Table 1.

In addition to the SMFs, we also incorporate the ratio of the passive
and SF SMFs into our analysis. The ratio takes into account that the
amplitudes of the passive and SF SMFs are correlated to some degree,
given that they are measured from the same sample of galaxies. All
measurements are made in three redshift bins that span a range of
$z=[0.22,1.00]$. The median redshifts are $z=0.36$, 0.66, and 0.88. We
will present our measurements in \S 4 when discussing our best-fit
models.

We also incorporate information from the COSMOS X-ray group catalog of
\cite{george_etal:11}. The central galaxy in each group is determined
with high probability, yielding a measurement of the red fraction of
central galaxies at $\mhalo \sim 10^{13.5} \msol$ in each redshift
bin. The error in this quantity is determined by bootstrap resampling
of the group catalog. The purpose of including these data is to
prevent unphysical divergent behavior of the models, e.g., models in
which the red fraction of central galaxies turns over and approaches
zero at high halo masses where the constraints from the three galaxy
measures are weak. In practice, the inclusion of the group data does
not significantly affect the results.

As described in L11 and L12, we use a large-volume, high-resolution
N-body simulation to create mock galaxy distributions with the same
angular size and comoving depth of each slice in the COSMOS survey. We
use the ``Consuelo'' simulation, which is part of the LasDamas
simulation suite (C. McBride, in preparation). This simulation is 420
\hmpc\ on a side and contains 1400$^3$ particles. These mocks are then
used to estimate the covariance matrices of each data set. Within the
simulation, we are able to create 409, 179, and 105 mocks for the
$z=0.36$, $0.66$, and $0.88$ redshift bins, respectively. We populate
the halos in the simulation with galaxies using a preliminary HOD fit
to the measurements, yielding a preliminary estimate of the
covariances. We then repeat this procedure with HOD fits that utilize
the first covariance matrices to produce the final errors used in the
results presented here. These covariance matrices are used as the full
errors on the SMFs and $\wtheta$ measurements. Because these two
statistics involve simple counting of galaxies and their pairs, the
N-body simulations encompass both the sample variance from large-scale
structure and shot noise from small number statistics. For the lensing
measurements, the statistical errors arising from the ellipticity
measurements of the background sources are added to the covariance
matrices from the mocks, which estimate the sample variance. In most
cases, the statistical errors dominate the uncertainty in $\DS$.

% **************** TABLE 3 ***************
% MCMC constraints
\begin{deluxetable}{cccc}
\tabletypesize{\scriptsize} 
\tablecolumns{4} 
\tablewidth{20pc} 
\tablecaption{HOD values from MCMC\label{t.hod_results}} 
\tablehead{\colhead{Parameter} &\colhead{$z_1$}
  &\colhead{$z_2$} & \colhead{$z_3$} }
\startdata
active galaxies & & & \\ \hline \vspace{0.0cm}\\  

$\log M_{1}$ &           $12.56 \pm 0.05$ & $12.77 \pm 0.05$ & $12.69 \pm 0.04$ \\            
$\log M_{\ast,0}$ &      $10.96 \pm 0.06$ & $10.98 \pm 0.03$ & $10.97 \pm 0.02$ \\            
$\beta$ &                $0.44 \pm 0.02$ & $0.46 \pm 0.03$ & $0.43 \pm 0.02$ \\               
$\delta$ &               $0.52 \pm 0.29$ & $1.15 \pm 0.31$ & $0.73 \pm 0.25$ \\               
$\gamma$ &               $1.48 \pm 0.43$ & $2.15 \pm 0.51$ & $4.71 \pm 0.56$ \\               
$\sigsm$ &               $0.21 \pm 0.06$ & $0.24 \pm 0.02$ & $0.25 \pm 0.01$ \\               
$\Bcut$ &                $0.28 \pm 1.91$ & $0.22 \pm 1.09$ & $0.18 \pm 1.10$ \\               
$\Bsat$ &                $33.96 \pm 19.61$ & $24.55 \pm 21.29$ & $112.70 \pm 26.81$ \\        
$\bcut$ &                $0.77 \pm 1.80$ & $0.62 \pm 0.96$ & $1.00 \pm 1.42$ \\               
$\bsat$ &                $1.05 \pm 0.44$ & $1.16 \pm 0.52$ & $2.65 \pm 0.39$ \\               
$\asat$ &                $0.99 \pm 0.20$ & $0.96 \pm 0.18$ & $0.84 \pm 0.14$ \\               
\vspace{0.05cm} \\\hline
passive galaxies & & & \\ \hline \vspace{0.0cm}\\  

$\log M_{1}$ &           $12.08 \pm 0.20$ & $12.18 \pm 0.23$ & $12.21 \pm 0.17$ \\            
$\log M_{\ast,0}$ &      $10.70 \pm 0.10$ & $10.78 \pm 0.13$ & $10.83 \pm 0.10$ \\            
$\beta$ &                $0.32 \pm 0.09$ & $0.13 \pm 0.07$ & $0.02 \pm 0.04$ \\               
$\delta$ &               $0.93 \pm 0.25$ & $0.81 \pm 0.18$ & $0.44 \pm 0.11$ \\               
$\gamma$ &               $0.81 \pm 0.58$ & $0.09 \pm 0.41$ & $0.81 \pm 0.23$ \\               
$\sigsm$ &               $0.28 \pm 0.03$ & $0.21 \pm 0.03$ & $0.18 \pm 0.05$ \\               
$\Bcut$ &                $21.42 \pm 10.34$ & $0.01 \pm 0.01$ & $0.21 \pm 1.42$ \\             
$\Bsat$ &                $17.90 \pm 22.99$ & $21.35 \pm 9.50$ & $13.16 \pm 3.83$ \\           
$\bcut$ &                $-0.12 \pm 0.46$ & $-1.55 \pm 1.53$ & $0.46 \pm 0.80$ \\             
$\bsat$ &                $0.62 \pm 0.52$ & $0.58 \pm 0.14$ & $0.77 \pm 0.22$ \\               
$\asat$ &                $1.08 \pm 0.26$ & $1.15 \pm 0.10$ & $0.98 \pm 0.12$ \\               
\vspace{0.05cm} \\\hline
passive central fraction & & & \\ \hline \vspace{0.0cm}\\  
$\log \fred(M_1)$ &      $-1.28 \pm 0.20$ & $-7.32 \pm 2.32$ & $-6.89 \pm 2.18$ \\            
$\log \fred(M_2)$ &      $-0.85 \pm 0.10$ & $-1.17 \pm 1.04$ & $-1.23 \pm 1.47$ \\            
$\fred(M_3)$ &           $0.54 \pm 0.07$ & $0.47 \pm 0.10$ & $0.43 \pm 0.09$ \\               
$\fred(M_4)$ &           $0.63 \pm 0.05$ & $0.68 \pm 0.07$ & $0.59 \pm 0.08$ \\               
$\fred(M_5)$ &           $0.77 \pm 1.36$ & $0.81 \pm 0.20$ & $0.76 \pm 0.17$ \\

\enddata
\end{deluxetable}

\section{Theory}

In L11, we outlined an HOD-based model that can be used to
analytically predict the SMF, g-g lensing, and clustering signals. A
key component of this model is the SHMR which is modelled as a
mean-log relation, noted as $\mgal=\fshmr(M_h)$, with a
log-normal scatter\footnote{Scatter is quoted as the standard
  deviation of the logarithm base 10 of the stellar mass at fixed halo
  mass.} noted $\sigma_{\rm log M_{\ast}}$. Here we give a brief review
of the model of the minor modifications used to adapt it to passive and
SF subsamples of galaxies.

\subsection{The stellar-to-halo mass relation for central galaxies}

Following \citet{behroozi_etal:10}, $\fshmr(M_h)$ is
mathematically defined following its inverse function:

\begin{displaymath}
 \log_{10}(\fshmr^{-1}(\mgal)) = \hspace{0.65\columnwidth}
 \end{displaymath}
 \vspace{-3ex}
\begin{equation}
 \label{e.shmr}
\quad \log_{10}(M_1) + \beta\,\log_{10}\left(\frac{\mgal}{M_{\ast,0}}\right) +
 \frac{\left(\frac{\mgal}{M_{\ast,0}}\right)^\delta}{1 + \left(\frac{M_{\ast}}{M_{\ast,0}}\right)^{-\gamma}} - \frac{1}{2}.
\end{equation}

\noindent where $M_{1}$ is a characteristic halo mass, $M_{*,0}$ is a
characteristic stellar mass, $\beta$ is the low-mass slope, and
$\delta$ and $\gamma$ control the massive end slope. We note that
equation \ref{e.shmr} is only relevant for central galaxies. We use
equation \ref{e.shmr} to parameterize the SHMR of both passive and SF
central galaxies, but each subsample will have a separate $\fshmr$.

Eq.~\ref{e.shmr} specifies the mean halo mass as a function of
$\mgal$. We assume that the distribution of central galaxy mass at
fixed halo mass, $\Phi_c(\mgal|\mhalo)$, follows a log-normal
distribution with scatter $\sigsm$. We will discuss halo occupation of
central galaxies at fixed halo mass presently. Previous work suggests
that $\sigsm$ is independent of halo mass. \cite{more_etal:11} finds a
scatter in $\mgal$ at fixed halo mass of $0.17 \pm 0.04$
dex. \cite{moster_etal:09} are able to fit the SDSS galaxy clustering
measurements assuming constant $\sigsm$. In L12 we found that a
halo mass-varying scatter produced no better fit than a model with
constant scatter. We adopt a constant $\sigsm$ here as well, but allow
the scatter for passive and SF central galaxies to be independent.

\subsection{Accounting for passive and star-forming subsamples}

We are bound by the requirement that each halo contains one and only
one central galaxy. The mass of that galaxy may be too small to be
counted in any COSMOS sample, but formally we require that

\begin{displaymath}
\int \fred(\mhalo)\times\Phi_{\rm cen}^{\rm q}(\mgal|\mhalo)\,+ \hspace{0.25\columnwidth}
\end{displaymath}
\begin{equation}
[1-\fred(\mhalo)]\times\Phi_{\rm cen}^{\rm SF}(\mgal|\mhalo)\,d\mgal = 1,
\end{equation}

\noindent where $\fred(\mhalo)$ is a function specifying the fraction
of times that a halo of mass $\mhalo$ contains a quenched central
galaxy (independent of galaxy mass), and $\Phi_{\rm cen}^{\rm
  x}(\mgal|\mhalo)$ is the conditional stellar mass function for
central quenched or SF galaxies, each normalized to
unity. Parameterizing the quenching of central galaxies by halo mass
as opposed to stellar mass (or the ratio between the two) makes an
implicit choice of the mechanisms that quench star formation in
central galaxies (see the discussions in \citealt{hopkins_etal:08b}
and \citealt{tinker_wetzel:10}). Given the small scatter between
stellar mass and halo mass, this choice is not likely to bias the
results we focus on here, e.g., the fraction of centrals that are
red. This choice is also beneficial for its ease of implementation in
our halo occupation framework.

We do not choose a parametric form for $\fred(\mhalo)$. Rather, we
choose five halo mass points at which to specify $\fred(\mhalo)$ and
smoothly interpolate between them. The five masses are evenly spaced
in $\log\mhalo$ from 10.8 to 14.0. 

\subsection{Calculating halo occupation of centrals and satellites}

In order to avoid explicit dependence of our HOD parameters on our bin
size, we define all HODs as threshold quantities. Having halo
occupation parameterized for threshold samples yields maximal
flexibility for taking the same HOD parameters and calculating
$\langle N\rangle_M$ for a bin of arbitrary size. For a sample of
galaxies above a threshold stellar mass, the central occupation
function $\ncen$ is expressed as

\begin{displaymath}
  \langle N_{\rm cen}(M_h|>\mgal) \rangle = \hspace{0.65\columnwidth}
\end{displaymath}
\begin{equation}
\label{e.ncen}
\frac{1}{2}\left[ 1-\mbox{erf}\left(\frac{\log_{10}(\mgal) - \log_{10}(\fshmr(M_h)) }{\sqrt{2}\sigma_{\rm log M_{*}}} \right)\right].
\end{equation}

\noindent As discussed in L11, equation (\ref{e.ncen}) correctly
captures the behavior of $\ncen$ for massive galaxy samples, as
opposed to the common parameterization where scatter is parameterized
at fixed stellar mass as opposed to fixed halo mass.  Eq.~\ref{e.ncen}
is valid for both SF and passive central galaxies, but the parameters of the
$\fshmr$ are independent for each subsample. Eq.~\ref{e.ncen} assumes
that there is one central galaxy per halo; in the case of our
subsamples, this is not explicitly true. For red central galaxies,
Eq.~\ref{e.ncen} is multiplied by $\fred(\mhalo)$, and by
$1-\fred(\mhalo)$ for SF central galaxies.

The occupation of satellite galaxies  as a function of halo mass,
$\nsat$, is

\begin{displaymath}
\langle N_{\rm sat}(M_h|>\mgal) \rangle = \hspace{0.65\columnwidth}
\end{displaymath}
\begin{equation}
\label{e.nsat}
\left(\frac{\mhalo}{M_{\rm sat}}\right)^{\asat} \exp\left(\frac{-(\mcut+\fshmri(\mgal)}{\mhalo}\right),
\end{equation}

\noindent where $\msat$ is the halo mass scale for satellite galaxies,
$\mcut$ is a cutoff scale, and $\asat$ is how the number of satellites
scales with halo mass. We treat the satellite occupation of the passive
and SF subsamples independently; unlike central galaxies, there is
no integral constraint on the total number of satellite galaxies any
halo can have. Equation (\ref{e.nsat}) is a minor modification from
L11 (Eq. 12 therein); in L11, $\nsat$ is proportional to
$\ncen$---this guarantees that satellite occupation fully cuts off at
the same halo mass scale as central galaxies of the same
mass. However, in our new red/blue parameterization this would
correlate $\ncen$ to $\fred(\mhalo)$. We circumvent this problem by
including $\fshmri$ to the numerator in the exponential cutoff,
producing a similar cutoff scale.

HOD modeling of luminosity-dependent galaxy clustering has shown that
$\msat$ is roughly 20 times $\fshmri$, varying weakly with luminosity
(e.g., \citealt{zehavi_etal:05, zehavi_etal:11, zheng_etal:07,
  zheng_etal:09}). We thus parameterize $\msat$ and $\mcut$ as

\begin{equation}
\frac{M_{\rm sat}}{10^{12} M_{\odot}}= B_{\rm sat} \left(\frac{\fshmri}{10^{12} 
    M_{\odot}}\right)^{\beta_{\rm sat}},
\end{equation}

\noindent and 

\begin{equation}
\label{mcut_eq}
\frac{M_{\rm cut}}{10^{12} M_{\odot}}= B_{\rm cut} \left(\frac{\fshmri}{10^{12} 
    M_{\odot}}\right)^{\beta_{\rm cut}}.
\end{equation}

\noindent In L12 we set $\asat=1$, in agreement with many
previous results. However, the fraction of satellites that are star
forming depends on halo mass (\citealt{wetzel_etal:12_groups1}), thus
we allow $\asat$ to be free for both passive and SF
subsamples. Equations (\ref{e.ncen}) and (\ref{e.nsat}) give the
number of galaxies above a mass threshold as a function of halo
mass. Our data are measured in stellar mass bins. To determine the
halo occupation in a given bin, we simply take the difference between
$\nsat$ (or $\ncen$) at the low- and high-mass edges of the bin.

The model has 27 free parameters. To model the halo occupation of a
given subsample requires 11 free parameters. The SHMR has 5 free
parameters ($M_{1}, M_{*,0}, \beta, \delta, \gamma$), with one
additional parameter for the scatter, $\sigsm$. The satellite
occupation requires 5 more parameters ($\Bsat$, $\bsat$, $\Bcut$,
$\bcut$, $\asat$). To determine the fraction of central galaxies that
are red at each halo mass requires 5 more parameters for a total of
27. Each set of 27 parameters describes the galaxy-halo relation at a
given redshift. For each of our three redshift bins, we fit the
parameters separately. We use the halo mass function of
\cite{tinker_etal:08_mf}, the halo bias relation of
\cite{tinker_etal:10_bias}, and the concentration-mass relation for
dark matter halos of \cite{munoz_cuartas_etal:11}, assuming that
satellite galaxies follow the dark matter within a halo with an NFW
profile (\citealt{nfw:97}). We refer the reader to L11 for a complete
description of how to take the halo occupation parameters and
calculate the SMFs, clustering, and lensing signals.

\begin{figure*}%{figure}
%\epsscale{1.0} 
\plotone{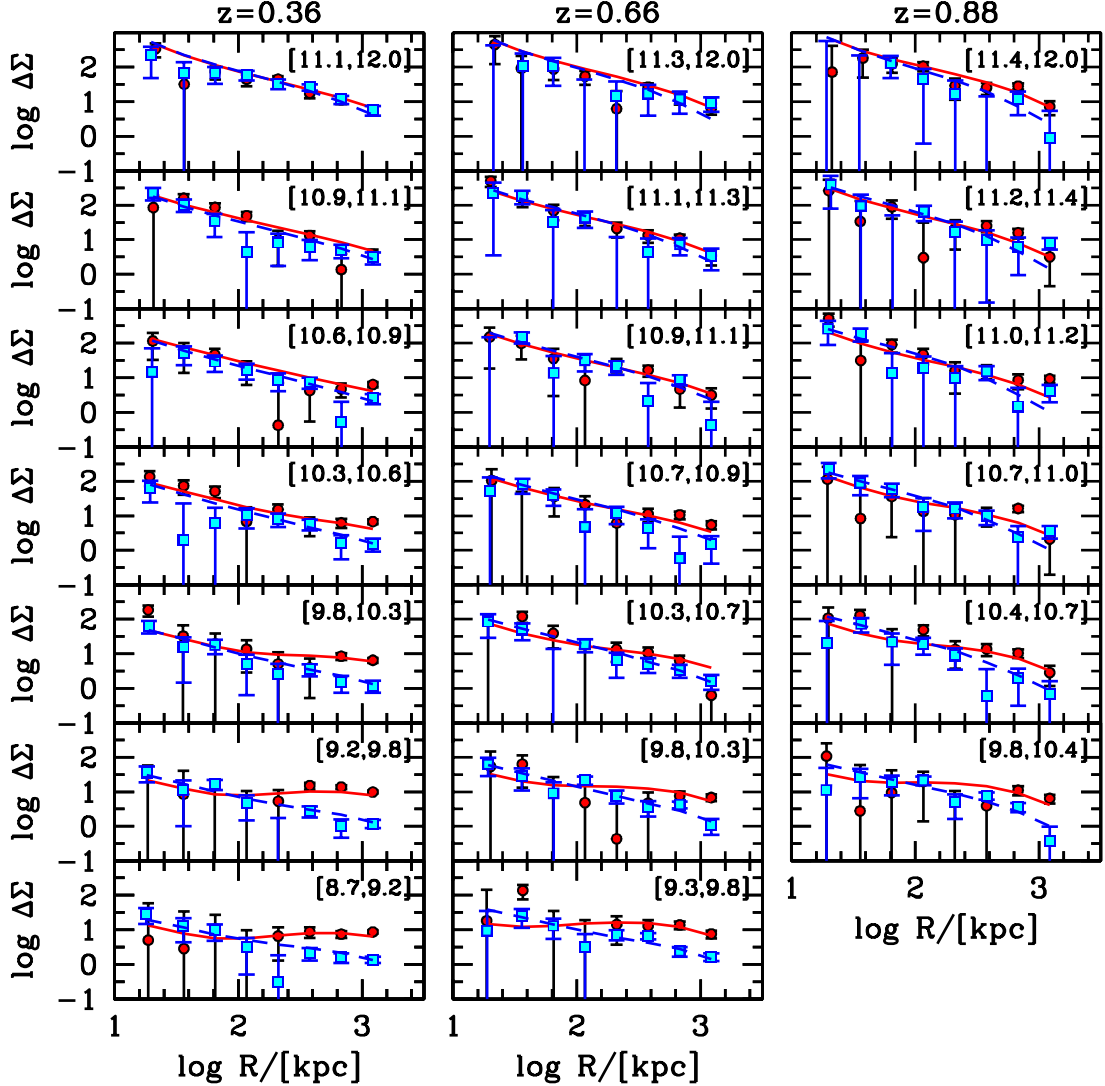}
%\vspace{-7.5cm}
\caption{ \label{bestfit_lensing} Galaxy-galaxy lensing of COSMOS
  galaxies in stellar mass bins. Point with error bars are
  measurements while curves indicate best-fit HOD models. Colors and
  point types are the same as Figure \ref{bestfit_smf}. Stellar mass
  bins for the lensing measurements can be found in Table 1.  A
  breakdown of the components of the fits for four examples can be
  found in Figure \ref{lensing_detail}. }
\end{figure*}%{figure}

\begin{figure}%{figure}
\epsscale{1.2} 
\plotone{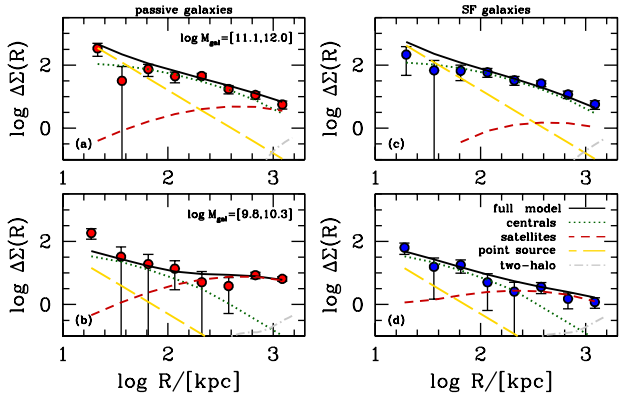}
\vspace{-3.3cm}
\caption{ \label{lensing_detail} A breakdown of the lensing fits for
  two stellar mass bins in the $z=0.36$ redshift bin. The top row
  (panels [a] and [c]) show the most massive galaxy bin for passive
  and SF galaxies. The bottom row shows the $[9.8,10.3]$ stellar mass
  bin. The solid curve (black) shows the overall fit, which is the sum
  of the other curves. The dotted curve (green) is the lensing profile
  of the dark matter halo around central galaxies. The short-dash
  curve (red) shows the lensing profile of the halos around satellite
  galaxies. The long-dash (yellow) curve represents he central
  point-source---i.e., the central galaxy itself. The dash-dot (gray)
  curve is the lensing contribution from nearby halos---i.e., the
  two-halo term.  For both mass bins, the passive galaxies have a
  higher fraction of satellites, evinced by the higher amplitude of
  the satellite lensing signal. At fixed $\mgal$, the halos that host
  the central galaxies are roughly equal mass between passive and SF
  galaxies (this does not mean that $\fshmr$ is the same---we will
  discuss the differences between $\langle \mgal|\mhalo \rangle$ and
  $\langle\mhalo|\mgal\rangle$ in the following section. }
\end{figure}%{figure}

%-------------------------------
% -- TABLE for chi^2 values
%-------------------------------
\begin{deluxetable}{ccc}
\tabletypesize{\scriptsize} 
\tablecolumns{3} 
\tablewidth{15pc} 
\tablecaption{$\chi^2$ values for best-fit models\label{t.chi2_values}} 
\tablehead{\colhead{$z=[0.22,0.48]$} & \colhead{$z=[0.48,0.74]$} & \colhead{$z=[0.74,1.00]$}}
\startdata
& & \\
$218.5/(247-27)$ & $273.0/(241-27)$ & $220.5/(207-27)$ \\
\enddata
\tablecomments{The value in each column is the $\chi^2$ value divided by
  the number of data points minus the number of free parameters.}
\end{deluxetable}

\section{Results}

We use Markov Chain Monte Carlo (MCMC) analysis to find both the
best-fit model and the uncertainties in model parameters. We analyze
each redshift bin separately. For each trial model in the MCMC chain,
we calculate a separate $\chi^2$ for the SMF, for each mass bin in
$\wtheta$, and each mass bin in $\DS$, for passive and SF subsamples,
and the red fraction of central galaxies within the X-ray groups. The
total $\chi^2$ is then

\begin{equation}
\label{e.chi2}
\chi^2_{\rm tot} = \sum_{q,SF}\left[ \chi^2_{\rm smf} +
  \sum_{i=1}^{N_w} \chi^2_{w,i} + \sum_{j=1}^{N_{\DS}}
  \chi^2_{\DS,j}\right] + \chi^2_{fred} + \chi^2_{ratio}.
\end{equation}

\noindent The last two terms in the above equation represent the
$\chi^2$ for the red central fraction from the X-ray group catalog and
the $\chi^2$ for the ratio of the passive ($q$) and SF SMFs,
respectively. We use a covariance matrix for each individual $\chi^2$
calculation, with the exception of $\chi^2_{fred}$. Parameter values
and errors from the MCMC chains are in Table \ref{t.hod_results}. The
total $\chi^2$ for each best-fit model is listed in Table
\ref{t.chi2_values}.

\subsection{Stellar Mass Functions and the Quenched Fraction of Galaxies}

Figure \ref{smf_compare} shows our measurements of the passive and SF
SMFs in COSMOS. Data are shown down to the stellar mass completeness
limits for each subtype. The stellar mass functions show limited
evolution across our redshift range with the exception of low-mass
passive galaxies: the abundance of these galaxies increases by a
factor of 2-3 depending on stellar mass. This trend has been shown in
a number of papers as a component of the ``downsizing'' of galaxy
formation. \cite{brinchmann_ellis:00} detected this trend in
morphologically-selected samples, and \cite{bundy_etal:06} found
similar results in the abundances of SF and passive galaxies in
DEEP2. In our measurements, the $z=0.36$ passive SMF shows a minimum
at $mgal\sim 10^{9.5} \msol$, with a subsequent upturn at lower
masses, as shown by \cite{drory_etal:09} for COSMOS data and confirmed
in PRIMUS by \cite{moustakas_etal:13}.

In Figure \ref{smf_compare} we compare our measurements to those from
\cite{drory_etal:09} and \cite{moustakas_etal:13}. The
\cite{drory_etal:09} measurements are also taken from COSMOS, but with
two main differences. First, they are measured in different redshift
bins. Due to the small footprint of COSMOS, the sample variance from
different binning is a non-negligible effect. Second, there are
differences in the stellar mass calculations themselves: Drory
et.~al.~fit the mass-to-light ratio $(M/L)$ from all photometric bands,
while the method of Bundy et.~al.~uses multi-band colors to constrain
the M/L ratio in the observed K-band, with mass estimates derived from
application to the K-band luminosity only. There are also minor
differences in the stellar population templates used.  Last, in this
figure we plot the fitting function results rather than the
measurements themselves. The Drory et.~al.~measurements lie slightly
above their fits at the massive end, so the agreement with our data is
somewhat better than implied in this figure. Even so, there are
minimal differences in the SMFs.

The SMFs from \cite{moustakas_etal:13} are measured from PRIMUS (their
Fig 11; tabulated data kindly provided by J. Moustakas), which is a
larger area but does not go as deep as COSMOS due to the use of
low-resolution spectroscopy to obtain galaxy redshifts. The abundance
of passive galaxies is somewhat higher in the PRIMUS results, but the
conclusion of \cite{moustakas_etal:13} agrees with our measurements
here: that the only significant change in abundance is in the low-mass
passive population.

Because the focus of this paper is on the growth of the red sequence,
we compare our measurements for the redshift evolution of the overall
quenched fraction to that of recent measurements from PRIMUS
(\citealt{moustakas_etal:13}) and to the analysis of zCOSMOS by
\cite{knobel_etal:13}. We define the quenched fraction as the density
of passive galaxies relative to the total number. PRIMUS contains
within it the COSMOS field, but there are differences in both the
stellar mass assignment and in the determination of which galaxies are
passive. \cite{moustakas_etal:13} use SED fitting to estimate the star
formation rates of PRIMUS objects and then divide the sample based
upon this distribution. Figure \ref{cosmos_primus_redfrac} compares
the quenched fractions for four different stellar masses between the
two surveys. At all masses, the PRIMUS $\fred$ is slightly higher than
the COSMOS value(s). An important comparison, however, is the slope of
$\fred$ with redshift. For each bin in $\mgal$, the rate of change
appears consistent between the two surveys. For zCOSMOS, the flux
limit makes it difficult to achieve a long redshift baseline for
anything but the most massive galaxies. But the quenched fractions in
\cite{knobel_etal:13} are significantly higher than either PRIMUS or
this work. \cite{knobel_etal:13} use a $U-B$ color cut to select their
sample of passive galaxies, which may be susceptible to dust
contamination. In their paper they compare their quenched fractions to
those derived from a $NUV-R-J$ color-color diagram (similar to the
approach used here), finding very good agreement. In contrast to their
results (their Figure 1), the single-color cut is not consistent with
the $NUV-R-J$ color selection and it yields a decreasing $\fred$ with
decreasing $z$ at the most massive galaxies, which is at odds with the
other two results. We will make further comparisons with the results
of \cite{knobel_etal:13} in the \S 4.5.

In this figure we have included data from the SDSS groups catalog of
\cite{tinker_etal:11_groups}.  These represent the data points at
$z=0.05$. In this figure, we are presenting $\fq$ for the overall
galaxy population, but the group finder is applied to volume-limited
samples derived from the SDSS Main sample, yielding a full
central-satellite decomposition of all galaxies in the sample. This
group catalog is $\sim 95\%$ complete in finding central galaxies and
$\sim 90\%$ pure in its sample of satellite galaxies. Quenched
fractions in sub-populations of the group catalog are corrected for
impurity and completeness statistically (see further details in
\citealt{tinker_etal:11_groups}). We will make significant use of this
catalog later in the paper. The differences in stellar mass estimates
between COSMOS and SDSS make comparisons of absolute abundances
problematic, but fractions are more robust. To facilitate a more
robust comparison of the SDSS data with our COSMOS results, we have
added 0.2 dex to the stellar mass estimates and added 0.2 dex of
scatter. The former represents the 0.2 dex shift in the SMFs between
SDSS (\citealt{li_white:09}) and COSMOS once deconvolved to a common
scatter value. The latter is meant to mock up the increase
uncertainties between SDSS spectroscopic redshifts and COSMOS
photometric redshifts (see the discussion in Figure 14 in L12).  Both
of these changes lower $\fq$ by 0.1 to 0.2, with the shift in mass
scale dominating the effect. The upper error bars show the original
SDSS values before shifting and adding scatter. Because both
alterations to the SDSS data lower $\fq$, the values used here should
be considered lower limits on the quenched fraction of SDSS galaxies.

\subsection{Comparison of the Measurements to the Best Model Fits}

Figure \ref{bestfit_smf} compares the stellar mass functions to the
best-fit halo occupation models from the MCMC chains. The overall
model SMF is shown with the thick solid curves, and the contribution
to the SMFs from satellite galaxies is shown with the thin curves.
The lower panels show the abundance ratio of SF and passive
galaxies. In these panels, the growth of the red sequence at low mass
is more evident.

Figure \ref{bestfit_wtheta} shows the clustering measurements for the
passive and SF galaxies. Consistent with previous measurements from
other redshifts and other surveys, the passive galaxies have equal or
higher clustering than the SF galaxies at every bin of stellar
mass. For low-mass galaxies, the enhanced clustering of passive
galaxies is due to the high fraction of such galaxies being satellites
in high-mass halos (e.g., \citealt{zehavi_etal:05,
  tinker_etal:08_voids, vdb_etal:03_early_late, skibba_sheth:09,
  weinmann_etal:06a, tinker_wetzel:10, wetzel_etal:12_groups1}) This
effect gives rise to the well-known color-density relation. At high
masses, $\log\mgal\ge 10.8$, the large-scale bias appears roughly
independent of color, while the small-scale clustering of passive
galaxies is slightly enhanced. As we will see when inspecting the
constraints on the SHMR in Figures \ref{shmr23}, massive SF galaxies
live in higher mass halos than their red counterparts, when binned by
halo mass; this is true of both the SHMR results presented here and the
group catalog results from \cite{tinker_etal:12_cosmos} (hereafter
T12). However, when binned by galaxy mass, scatter minimizes the
difference in the mean halo mass and thus the large-scale
bias. Massive SF galaxies have nearly negligible satellite fractions
in comparison to massive passive galaxies (at least at $z\ge 0.48$),
yielding a higher amplitude for the passive galaxy subsample at small
scales.

Figure \ref{bestfit_lensing} rounds out our presentation of the data
and model fits. Lower-mass star-forming galaxies primarily live as
central galaxies in lower mass halos, the lensing signal is weaker
than that of passive galaxies and thus has larger statistical errors. This
is reflected in the large error bars for the lower-mass SF
measurements. To better understand the information that the lensing
signal affords, Figure \ref{lensing_detail} shows a breakdown of the
constituent parts of the lensing fit for high mass and low mass
galaxies. Ignoring the contribution to the lensing signal between two
halos\footnote{The two-halo terms is included in all modeling, but has
  minimal impact on our results because we do not measure $\DS$ out
  past 1 Mpc.}, the lensing signal has three parts: the halo profile
around central galaxies, the halo profile around satellite galaxies,
and the central point source (i.e., the galaxy itself). For red
galaxies, the higher amplitude of the $\DS$ measurements at scales
$R\gtrsim 100$ kpc is indicative of the higher satellite fractions, as
this scale probes the mass profile of the dark matter halo in its
outskirts. Interior to this scale, the lensing signal is a measure of
the mass of dark matter halos around central galaxies. For both bins
in $\mgal$ shown, the mean halo mass of centrals appears roughly
consistent between passive and SF subsamples. The differences are
driven primarily by the fraction of galaxies that are satellites.

\subsection{The Stellar-to-Halo Mass Ratios and their Evolution}

\begin{figure*}
\epsscale{1.0} 
\plotone{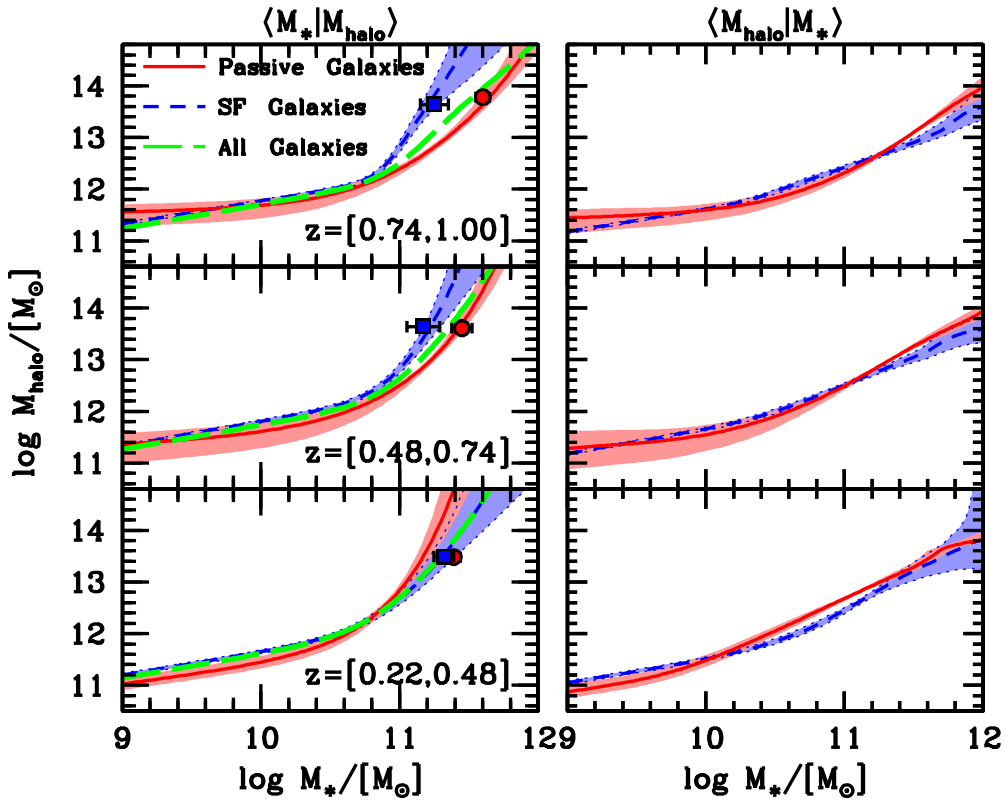}
\vspace{-4.0cm}
\caption{ \label{shmr23} Stellar-to-halo mass ratios for passive and SF
  central galaxies. Shaded regions indicate 68\% range in each
  quantity from the MCMC chains. The left-hand panels show $\fshmr$
  for each redshift bin, equivalent to the mean $\mgal$ at fixed
  $\mhalo$. The points with horizontal error bars represent the mean
  halo masses of the X-ray groups with passive and SF central
  galaxies, taken from T12. Long dashed curves show the SHMR for all
  galaxies, taken from L12. The right-hand panels show $\langle
  \mhalo|\mgal\rangle$. The larger scatter for SF galaxies creates
  more Eddington bias, thus when binned in $\mgal$, the mean halo mass
  is significantly smaller than $\fshmr$. Thus the lensing signals for
  massive galaxies are similar between passive and SF samples. }
\end{figure*}

The left-hand panels in Figure \ref{shmr23} show the SHMR for red and
SF galaxies at each redshift bin. The curves show the best-fit model
for each sample, while the shaded regions indicate the range that
contains inner 68\% of the models. At low masses, the SHMR becomes
shallow and stellar mass increases much more rapidly than halo mass:
$\mgal \sim \mhalo^{1/\beta}\sim \mhalo^2$. As galaxy mass increases,
however, the relation reaches a pivot point at which central galaxies
increase in mass slower than their halos and the SHMR becomes
steep. This is now accepted as a generic result of the abundance
matching paradigm (\citealt{conroy_etal:06, wang_etal:07,
  conroy_wechsler:09, moster_etal:09, behroozi_etal:10,
  yang_etal:11}). In L12 we defined the pivot point quantitatively as
the location in the $\mgal$-$\mhalo$ relation were the $\mgal/\mhalo$
ratio is maximal, usually around $\mgal\sim 10^{10}\,\msol$ and
$\mhalo\sim 10^{12}\,\msol$.

At all redshifts, the qualitative behavior of the SHMR for SF and
passive galaxies is quite similar; both subsamples show a pivot
point. The pivot halo mass is roughly $10^{12} \msol$ and the pivot
stellar mass is roughly $10^{10.6} \msol$. We will present a more
detailed comparison presently, but broadly speaking, there are few
major differences in the results. When comparing the results at low
masses, however, it is important to remember that these results do not
reflect the fraction of halos occupied by red central galaxies. For
the $z=0.66$ and $z=0.88$ redshift bins, the fraction of halos below
$10^{12} \msol$ that have red central galaxies is vanishingly
small. Only for $z=0.36$ does the red central fraction become
significant at these halo mass scales.

At scales above the pivot point, however, the behavior of the SHMR is
quantifiably different. At $z\sim 0.88$, massive SF galaxies occupy
larger halos at fixed stellar mass. In each panel, the point with horizontal
error bars shows the mean stellar mass within the X-ray group sample
from \cite{george_etal:11}. Although the red central fraction from the
groups is used within the MCMC chains, the mean stellar mass is not.
At $z\sim 0.66$, massive SF galaxies still reside in more massive
halos than their quiescent counterparts, but now the mean relations
are much closer together. At $z\sim 0.36$, the mean SMHR for red and
SF galaxies have crossed; massive passive galaxies occupy slightly more
massive halos than similar SF galaxies. The sample variance for the
low-$z$ bin is significant, but an evolutionary trend can be seen
across the full COSMOS sample. In T12 we compare these results to the
central galaxies found in the group catalog, finding quantitative
agreement. This figure also compares the new color-dependent results
to the SHMR from L12. At low masses, the SF SHMR tracks the
all-SHMR nearly exactly; this is expected given that SF galaxies
dominate the population at these masses. At high masses, the all-SHMR
is intermediate between the SF and passive SHMRs.

The origin of the differential evolution at the massive end comes from
our specific combination of data. The stellar mass functions clearly
indicate that there are more passive galaxies than SF galaxies at the
massive end of the spectrum. The clustering and lensing, however,
indicate that the large-scale bias and halo masses of the SF and passive
subsamples are consistent. Recall that the left-hand panels show the
mean stellar mass as a function of halo mass, even though we have
plotted the observable, $\log \mgal$, on the $x$-axis. At fixed
$\mgal$, scatter becomes very important at the massive end. The
right-hand panels in Figure \ref{shmr23} show the mean halo mass at
fixed stellar mass. In this plot, the differences between the red and
SF subsamples is almost entirely gone; thus, in bins of $\mgal$
where satellite galaxies are negligible (i.e., at stellar masses
significantly above the knee in the stellar mass function), one would
expect the clustering and lensing of SF and passive galaxies to be
consistent. The difference in the SHMRs is driven by the larger values
of $\sigsm$ for SF galaxies than for passive galaxies. For SF galaxies
at $z=0.88$, $\sigsm=0.25\pm 0.01$, while for passive galaxies
$\sigsm=0.18\pm 0.05$. By $z=0.36$, the passive galaxies have the smaller
scatter, and the steeper SHMR at the massive end. Although our
functional form for $\fshmr$ is meant to have a high degree of
flexibility at high halo masses, we cannot rule out a possible bias
due to our parametric form for $\fshmr$. Additionally, the assumption
of a symmetric, log-normal scatter may come into play in this regime
where the scatter is important. With the current data we are unable to
test alternative models for scatter.

\begin{figure}
\epsscale{1.6} 
\plotone{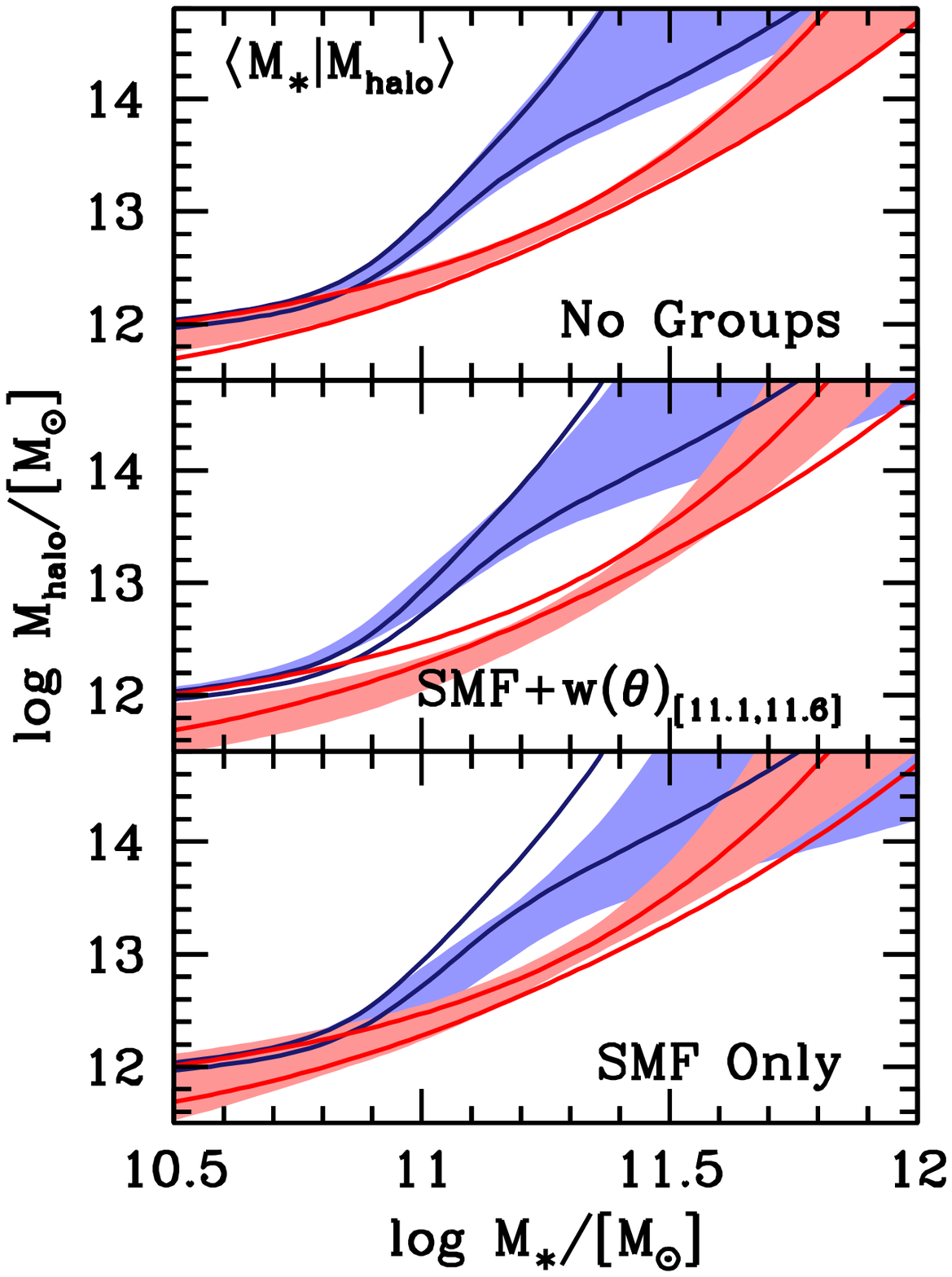}
\vspace{-1.0cm}
\caption{ \label{supp_chains} Stellar-to-halo mass ratios for passive and SF
  central galaxies when using only subsets of the available data. In
  each panel, the shaded region is the 68\% confidence interval for
  the SHMR from the MCMC chains. The lines indicate the same quantity
  from the original chains using all data (cf. Figure
  \ref{shmr23}). All panels show results from the $z=0.88$ redshift
  bin. {\it Bottom panel}: Chains that incorporate only the stellar
  mass functions and their ratio. {\it Middle panel}: The stellar mass
  functions, the SMF ratio, and the most massive clustering bin:
  galaxies with $\log\mgal=[11.1,11.6]$. {\it Top panel}: Chains using
  all data except $\fred(\mhalo)$ from the X-ray group catalog. }
\end{figure}

To determine the origin of the constraints on the high-mass end of the
SHMR, we ran a series of chains removing different data sets. Figure
\ref{supp_chains} show highlights from this series for the $z=0.88$
redshift bin. Intriguingly, the constraints when using the SMFs only
already show clear indication of a separation between the SHMR of SF
and passive galaxies, although the difference is not as large as the final
result. Adding just the most massive clustering bin increases the
separation between passive and SF SHMR values into rough agreement with
the full data. Similar results are found when {\it removing} the most
massive clustering bin and incorporating all others; constraints on
$\sigsm$ come from a range of stellar masses, provided the halos
occupied are in the regime where halo bias is monotonically
increasing with halo mass (roughly $\mhalo\sim 2\times 10^{11} \msol$
at this redshift). Because the halo bias function is highly
non-linear, the mean halo mass is not the same as the bias-weighted
halo mass. In this respect, the clustering has more constraining power
on $\sigsm$ than the lensing data. The top panel in Figure
\ref{supp_chains} demonstrates that our final results are not
sensitive to the data derived from the X-ray groups. Results when
removing the lensing data are similar.

\subsection{Central Red Fraction vs Halo Mass}

\begin{figure}
\epsscale{1.2} 
\plotone{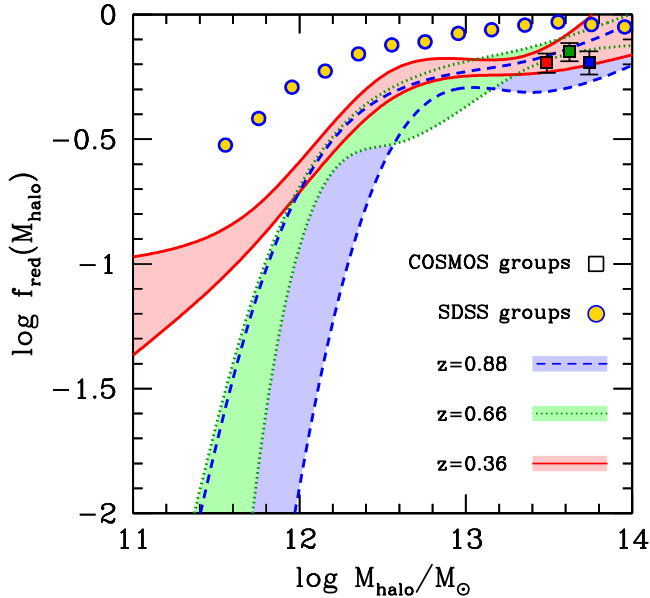}
\vspace{-0.8cm}
\caption{ \label{fred_halomass} The fraction of central galaxies that
  are red as a function of halo mass, $\fred(\mhalo)$, for all three
  redshift slices. The shaded regions show the 68\% range of values
  within the MCMC chains. The filled circles show the same quantity
  for the SDSS groups catalog of \cite{tinker_etal:11_groups}. The
  filled squares show the quenched fraction of central galaxies in
  COSMOS groups (\citealt{george_etal:11}) that are used in the MCMC
  modeling. At $z\ge 0.66$, there is a sharp cutoff in
  $\fred(\mhalo)$, implying that nearly all central galaxies at
  $\mgal\lesssim 10^{10.5}\,\msol$ are star forming at these
  redshifts. At lower redshifts, this cutoff moves to lower halo
  masses and there is a non-negligible contribution to the red
  sequence from low-mass central galaxies.  }
\end{figure}

Figure \ref{fred_halomass} shows the 68\% ranges of $\fred(\mhalo)$
from the MCMC chains for each redshift bin. At $z=0.66$ and $z=0.88$,
$\fred(\mhalo)$ has a sharp cutoff between $\mhalo = 10^{11.5-12.0}
\msol$. Although the median value of the cutoff evolves to somewhat
lower mass between $z=0.88$ and $z=0.66$, the results from the two
redshift bins are also consistent with no evolution. At $z=0.36$,
$\fred(\mhalo)$ is higher at all halo masses, most notably at $\mhalo
\lesssim 10^{11.5} \msol$; rather than a sharp cutoff in the quenched
central fraction, there is a long tail toward lower masses where the
$\fred(\mhalo)$ is 3-10\%. This is driven by all three sets of data: a
higher abundance of low-mass passive galaxies in the SMF, lower clustering
amplitude at for low-mass samples in $\wtheta$ and a lower satellite
fraction in the $\DS$ measurements. We will explore this in detail in
subsequent sections.

Figure \ref{fred_halomass} also shows results from the $z=0$ SDSS
groups catalog of \cite{tinker_etal:11_groups}. The shape of
$\fred(\mhalo)$ from the groups is similar to our non-parametric fit
in COSMOS, but the amplitude is higher by $\sim 0.1-0.2$ dex. This may
reflect evolution given that the time elapsed between $z=0.36$ and
$z=0.05$ is 3.3 Gyr, equal to the time elapsed from $z=0.88$ to
$z=0.36$. It may also reflect differences in the definition of
``quenched''; in \cite{tinker_etal:11_groups}, a 4000-\AA\ break below
1.6 is used to denote quenched, as opposed to NUV-optial-NIR colors
cuts used on the COSMOS data. Although this definition is less
sensitive to dust than the traditional $g-r$ color, $D_n4000$ may
suffer from aperture bias for more massive galaxies. The results for
COSMOS groups are plotted as well, one datum per redshift bin,
color-coordinated with the MCMC results.

\cite{tinker_wetzel:10} constrained halo occupation for color-selected
clustering from DEEP2 and COMBO17, concluding that there was not a
strong cutoff in $\fred(\mhalo)$ (additional data from the UKIDSS-UDS
were inconclusive). The clustering samples were created using a single
color cut without any NIR data, contaminating the red sequence with
dust-reddened star-forming galaxies. From Figure \ref{fred_halomass},
many of these galaxies are centrals in low-mass halos, making
$\fred(\mhalo)$ appear flatter and without any strong
cutoff. \cite{zhu_etal:11} find that $\sim 25\%$ of sub-$L^\ast$
galaxies with red colors are star-forming with specific rates of $\sim
10^{-10} {\rm yr}^{-1}$.

We note again that the detailed constraints on $\fred(\mhalo)$ depend
on our assumption that quenching of central galaxies is a
function of halo mass, independent of stellar mass. Because the mean
galaxy mass at fixed halo mass is similar between passive and SF
subsamples, a parameterization of $\fred$ that depends on stellar mass
rather than halo mass will likely yield consistent results.

\subsection{Central-Satellite Decomposition of the Stellar Mass Functions}

\begin{figure*}%{figure}
%\epsscale{1.3} 
\plotone{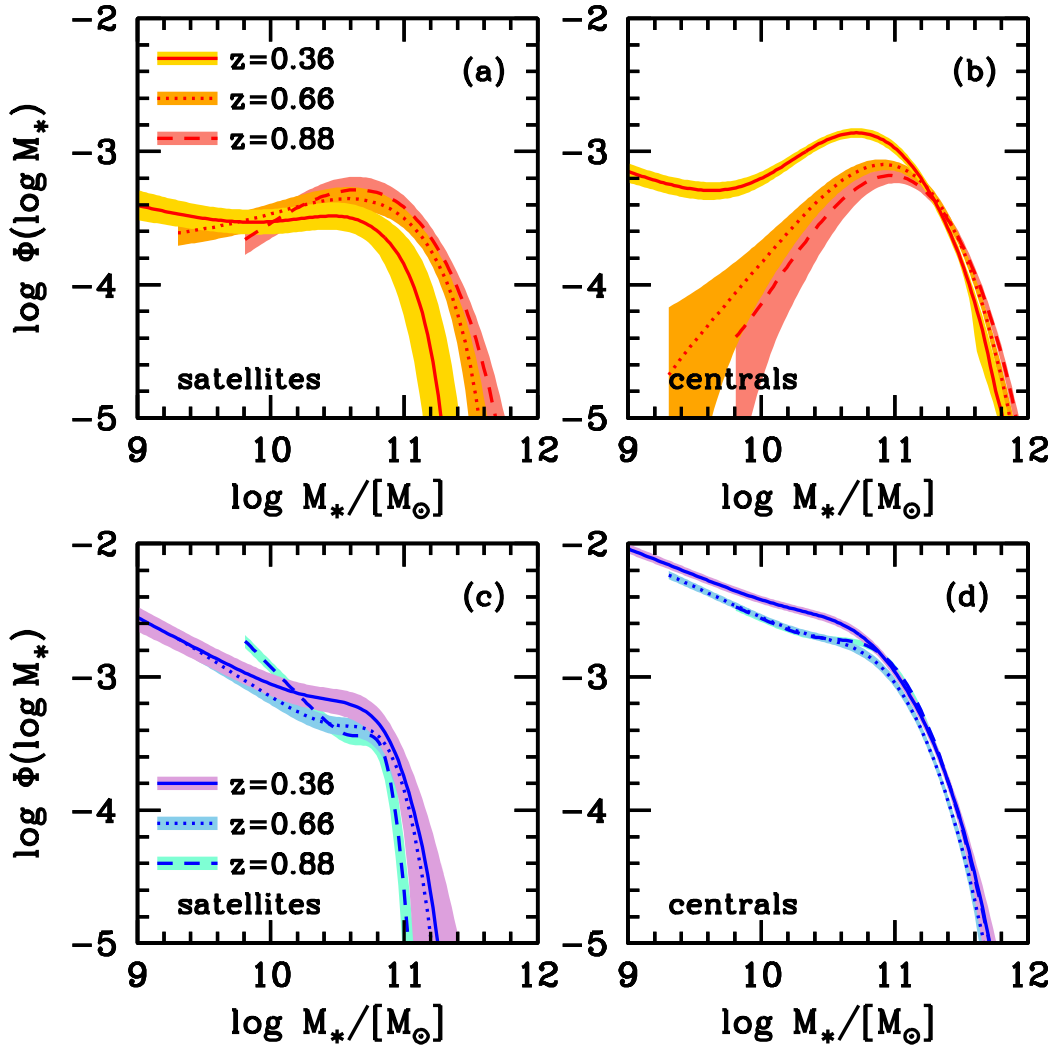}
%\vspace{-7.5cm}
\caption{ \label{smf_censat} The stellar mass functions of passive and
  SF cosmos galaxies broken down into the contributions from central
  and satellite galaxies. Panels (a) and (b) show results for quenched
  satellite and central galaxies, respectively. Panels (c) and (d)
  show results for star-forming satellite and central galaxies,
  respectively. The shaded regions represent the 68\% range of values
  within the MCMC chains. For all four subsamples, there is little
  redshift evolution at the high-mass end ($\mgal\gtrsim
  10^{11}\,\msol$). There is a dearth of high-mass quenched satellites
  at $z=0.36$, but this is likely a statistical outlier. At low
  masses, the only subsample that shows significant evolution is
  passive central galaxies; at $\mgal=10^{10}\,\msol$, the abundance
  of passive red centrals increases by more than an order of magnitude
  across our redshift baseline. Figure \ref{fred_evolve} shows that
  this growth in the fraction of quenched central galaxies continues to
  increases to $z=0$.}
\end{figure*}

Figure \ref{smf_censat} shows the SMFs for the SF and passive
subsamples, broken down into the separate abundances of central and
satellite galaxies. For SF galaxies, there is a modest increase in
the number of both central and satellite galaxies in time. The
abundance of red satellite galaxies exhibits little redshift evolution
at low masses. There is actually a deficit of massive red satellites
at $z=0.36$. It is unclear whether this represents physical evolution
versus sample variance, an issue we will discuss in this subsection.

The only subsample that exhibits significant redshift evolution is red
central galaxies. At the massive end there is minimal evolution,
consistent with larger surveys results of the evolution of the
luminosity function of Luminous Passive Galaxies (LRGs;
\citealt{cool_etal:08, wake_etal:06}). However, at $\mgal\lesssim
10^{11} \msol$, the number of red central galaxies increases rapidly
from $z=0.88$ to $z=0.36$. At $\mgal=10^{10} \msol$, this abundance
increases by 1.2 dex. At this same mass scale, the change in the
number of satellite galaxies is negligible.

\begin{figure*}%{figure}
%\epsscale{1.0} 
\plotone{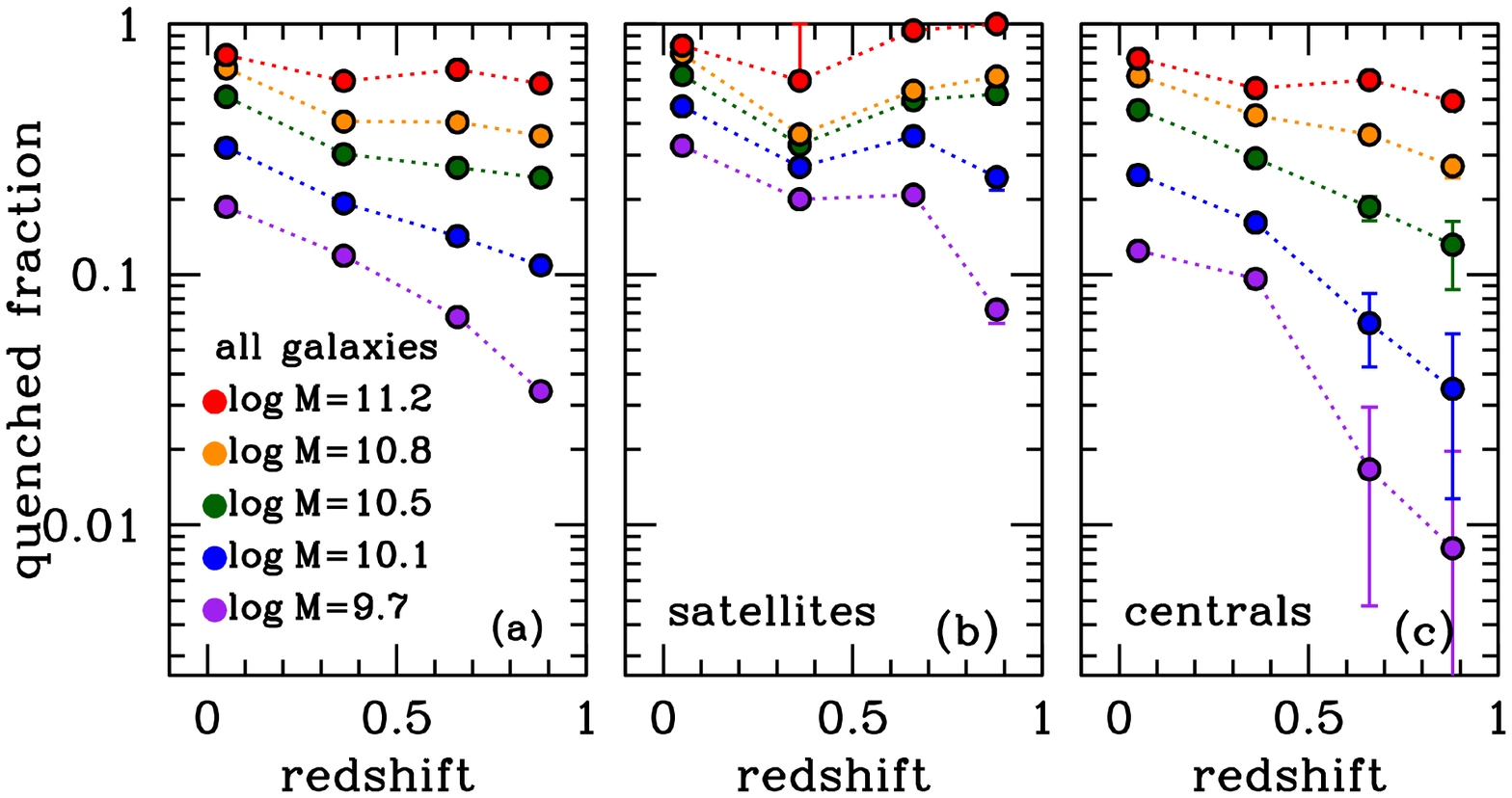}
\vspace{-6.5cm}
\caption{ \label{fred_evolve} The red (quenched) fraction of galaxies
  as a function of redshift for various stellar mass bins. Panel (a)
  shows $\fred$ for all galaxies. Panel (b) shows $\fred$ for
  satellite galaxies. Panel (c) shows $\fred$ for central
  galaxies. Error bars on the COSMOS measurements represent the 68\%
  range within the MCMC chains. Data points at $z=0.05$ are from the
  SDSS groups catalog of \cite{tinker_etal:11_groups}. The SDSS
  stellar masses have been modified to afford better comparison to
  COSMOS stellar masses, but these changes yield little to no change
  in the values on the $y$-axis. See text for details. Although the
  $z=0.36$ redshift bin is somewhat anomalous in its statistics, it is
  consistent with the overall trends in this figure: namely, the
  monotonic growth of the quenched fraction of galaxies at all masses,
  the near-constant quenched fraction of satellite galaxies, and the
  rapid growth of a population of quenched central galaxies,
  especially at low masses. }
\end{figure*}

This result is more clearly expressed by looking at the fraction of
galaxies that are red, and how this fraction depends on categorization
as a central or a satellite galaxy. Figure \ref{fred_evolve} shows
$\fred$ as a function of redshift for five values of $\mgal$ over the
range $\log\mgal=[9.7,11.2]$. In this figure we have included data
from the SDSS groups catalog of \cite{tinker_etal:11_groups}. The
differences in stellar mass estimates between the two surveys make
comparisons of absolute abundances problematic, but fractions are more
robust. To create the SDSS data in this figure, we have added 0.2 dex
to the stellar mass estimates and added 0.2 dex of scatter. The former
represents the 0.2 dex shift in the SMFs between SDSS
(\citealt{li_white:09}) and COSMOS once deconvolved to a common
scatter value. The latter is meant to mock up the increase
uncertainties between SDSS spectroscopic redshifts and COSMOS
photometric redshifts (see the discussion in Figure 14 in L12).
Both of these changes lower $\fq$ by 0.1 to 0.2, which the shift in
mass scale dominating the effect. The upper error bars show the
original SDSS values before shifting and adding scatter. Because both
alterations to the SDSS data lower $\fq$, the values
used here should be considered lower limits on the quenched fraction
of SDSS galaxies.

Figure \ref{fred_evolve}a shows $\fred$ for all galaxies in each
stellar mass bin. The rate of change in $\fred$ with redshift
monotonically decreases with increasing stellar mass. For massive
galaxies, $\fred$ is roughly constant. At $\log\mgal=9.7$, $\fred$
increases by a factor of five. Figure \ref{fred_evolve}b shows the
same quantity, but now for satellite galaxies only. Aside from the
lowest mass bin, $\fred(sat)$ in all bins is consistent with no
redshift evolution. Central galaxies, on the other hand, show
significant evolution; at $\log\mgal\lesssim 10$, $\fred(cen)$
increases by an order of magnitude. Even at $log\mgal=10.5$,
$\fred(cen)$ increases by a factor of 5 over our redshift baseline.

\begin{figure}
\epsscale{1.7}
\plotone{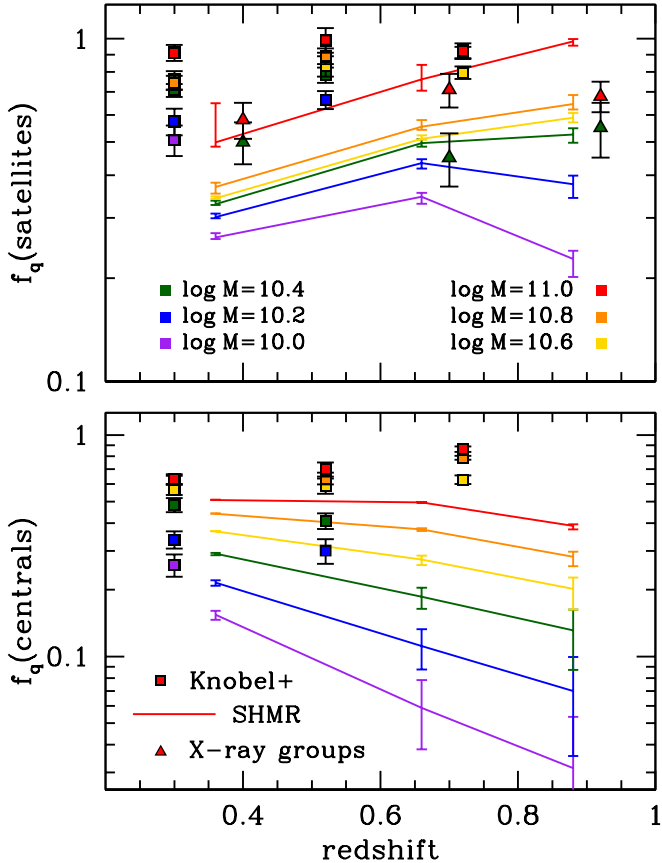}
%\vspace{-5.5cm}
\caption{\label{fred_evolve_knobel} The evolution of the quenched
  fraction from our SHMR analysis compared to that from the zCOSMOS
  groups catalog of \cite{knobel_etal:13}. The color scheme is the
  same as previous figures, but here we make the comparison in the
  stellar mass bins used in \cite{knobel_etal:13}. The top panel shows
  $\fq$ for satellites while the bottom panel shows $\fq$ for
  centrals. In the top panel, we also include results from the COSMOS
  X-ray group catalog of \cite{george_etal:13}. There is an overall
  shift in the total quenched fractions for the COSMOS and zCOSMOS
  samples (c.f., Figure \ref{cosmos_primus_redfrac}) such that the
  zCOSMOS sample has a higher fraction of quenched galaxies. For
  massive objects, the zCOSMOS sample has a decreasing $\fred$ with
  decreasing $z$, which is driven by the decrease in $\fred(cen)$ for
  bins at $\mgal>10^{10.5}\,\msol$. The photometric COSMOS sample used
  here has a monotonically increasing (or constant) overall red
  fraction, also driven by the behavior of the central galaxies. }
\end{figure}

\cite{knobel_etal:13} use group catalogs in the zCOSMOS survey to
measure the redshift evolution of centrals and satellites as well. Due
to the flux limit of the zCOSMOS target selection, they only achieve a
redshift baseline for galaxies $\mgal\gtrsim 10^{10.3}\,\msol$. They
also find little to no evolution of the red fraction of
satellites. For central galaxies, however, they find weaker evolution
for the red fraction of central galaxies. In Figure
\ref{fred_evolve_knobel} we compare $\fred$ for centrals and
satellites between the two methods. An objective comparison is
obstructed by the overall offset in $\fred$ between the two
definitions of quenched (see Figure \ref{cosmos_primus_redfrac}). Both
approaches yield a small decrease in $\fred(sat)$ for massive galaxies
as $z$ decreases, but the \cite{knobel_etal:13} groups yield a
quenched fraction at $\mgal=10^{11}\,\msol$ that is nearly unity. In
this panel we plot the results from the COSMOS X-ray groups of
\cite{george_etal:11}, which use the same definition of quenched as
this work. The $\fred(sat)$ values are mostly consistent with those
from our SHMR analysis. For central galaxies, the
\cite{knobel_etal:13} groups yield contrasting results above and below
$\mgal=10^{10.5}\,\msol$. Below this limit, the zCOSMOS central
galaxies show a moderate increase in $\fred(cen)$, but above this
limit the zCOSMOS central galaxies exhibit significantly decreasing
$\fred(cen)$ with decreasing redshift. The quenched fraction of
$\mgal=10^{11}\,\msol$ centrals decreases from 90\% to 60\% over their
redshift baseline.

The $U-B$ color cut used in zCOSMOS may be susceptible to dust
contamination, which may be stronger at higher redshifts where star
formation rates are also higher. Additionally, there may be differences
driven by the two methods---halo occupation and group
finding. Misclassification of which galaxy in a group is the central is
a major source of bias for group catalogs
(\citealt{skibba_etal:11}). Given that the quenched fraction of
satellites exhibits no redshift evolution, this type of bias will only
weaken the true trend of $\fred(cen)$. Moreover, \cite{knobel_etal:13}
use a probabilistic scheme to select subsamples of centrals and
satellite galaxies that have purity near 80\%, forcing them to assume
that these subsets are representative of the overall populations. Halo
occupation methods do not suffer from these biases, as central and
satellite populations are constrained only in a statistical fashion,
and not on an object-by-object basis. We also note that the central
galaxies in the X-ray group catalog used here are a much cleaner
sample of central galaxies, given that the group center can be
verified with the X-ray brightness profile.

\begin{figure*}%{figure}
%\epsscale{1.0} 
\plotone{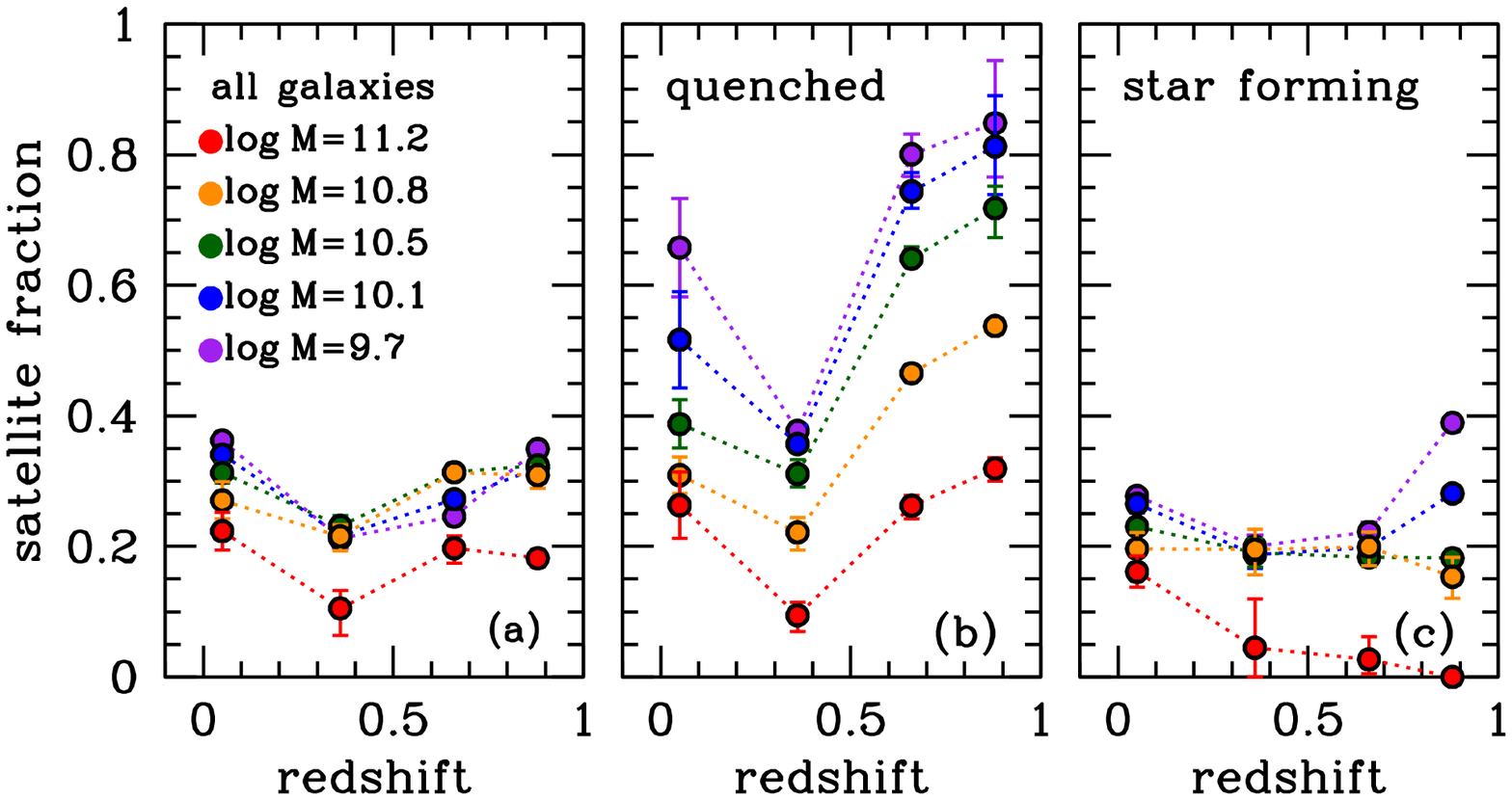}
\vspace{-6.5cm}
\caption{ \label{fsat_evolve} The satellite fraction of galaxies as a
  function of redshift for various stellar mass bins. Panel (a) shows
  $\fsat$ for all galaxies. Panel (b) shows $\fsat$ for red
  galaxies. Panel (c) shows $\fsat$ for star-forming galaxies. Error
  bars on the COSMOS measurements represent the 68\% range within the
  MCMC chains. Data points at $z=0.05$ are from the SDSS groups
  catalog of \cite{tinker_etal:11_groups}.The SDSS stellar masses have
  been modified to afford better comparison to COSMOS stellar masses,
  but these changes yield little to no change in the values on the
  $y$-axis. See text for details.  }
\end{figure*}

Figure \ref{fsat_evolve} shows a complementary statistic: the fraction
of galaxies that are satellites, $\fsat$, for the same stellar mass
bins and redshift range. For all galaxies, $\fsat$ is between
$0.25-0.35$, consistent with previous analyses of $z=0$
luminosity-dependent clustering (e.g., \citealt{zehavi_etal:05,
  tinker_etal:07, vdb_etal:07, zheng_etal:07, zehavi_etal:11}). Halo
occupation analysis of $z\sim 1$ luminosity dependent clustering
indicates a somewhat smaller $\fsat$ than at $z=0$
(\citealt{zheng_etal:07, abbas_etal:10}). However, recent analysis of
stellar-mass dependent clustering at $z=1-2$ by \cite{wake_etal:11}
find $\fsat$ values consistent with those in COSMOS. Due to the fact
that satellite galaxies are predominantly red, they are fainter than
SF galaxies at the same stellar mass, lowering the satellite
fraction at a given mass. For SF galaxies, star formation rates
increase with redshift, increasing the difference between luminosity
and stellar mass defined samples.

The satellite fractions of star forming galaxies are lower than for
the full sample, generally near $\sim 0.2$, with minimal redshift
evolution. Satellites dominate the population of low-mass passive galaxies
at $z\sim 1$. Even at $\log\mgal=10.8$, $\fsat=0.55$. By $z=0$,
satellites represent less than half of passive galaxies at
$\log\mgal>9.7$. The change in $\fsat$ for passive galaxies is
non-monotonic when incorporating the SDSS data, yielding a ``dip'' in
$\fsat$ at $z=0.36$. The small volume of this redshift slice raises
the possibility that the galaxy distribution around $z=0.36$ within
COSMOS is a significant outlier with respect to the cosmic mean. We
note that while the trend of $\fsat(red)$ with redshift is
non-monotonic, the trend in $\fred(cen)$ {\it is} monotonic. Thus, if
the $z=0.36$ redshift slice is simply removed from consideration, the
results in Figures \ref{fred_evolve} and \ref{fsat_evolve} are still
consistent with the scenario in which the only population to undergo
significant evolution since $z=1$ is red central galaxies.

\subsection{Signature of the Evolving Red Central Population in the
  Data}

\begin{figure*}%{figure}
\epsscale{1.1} 
\plotone{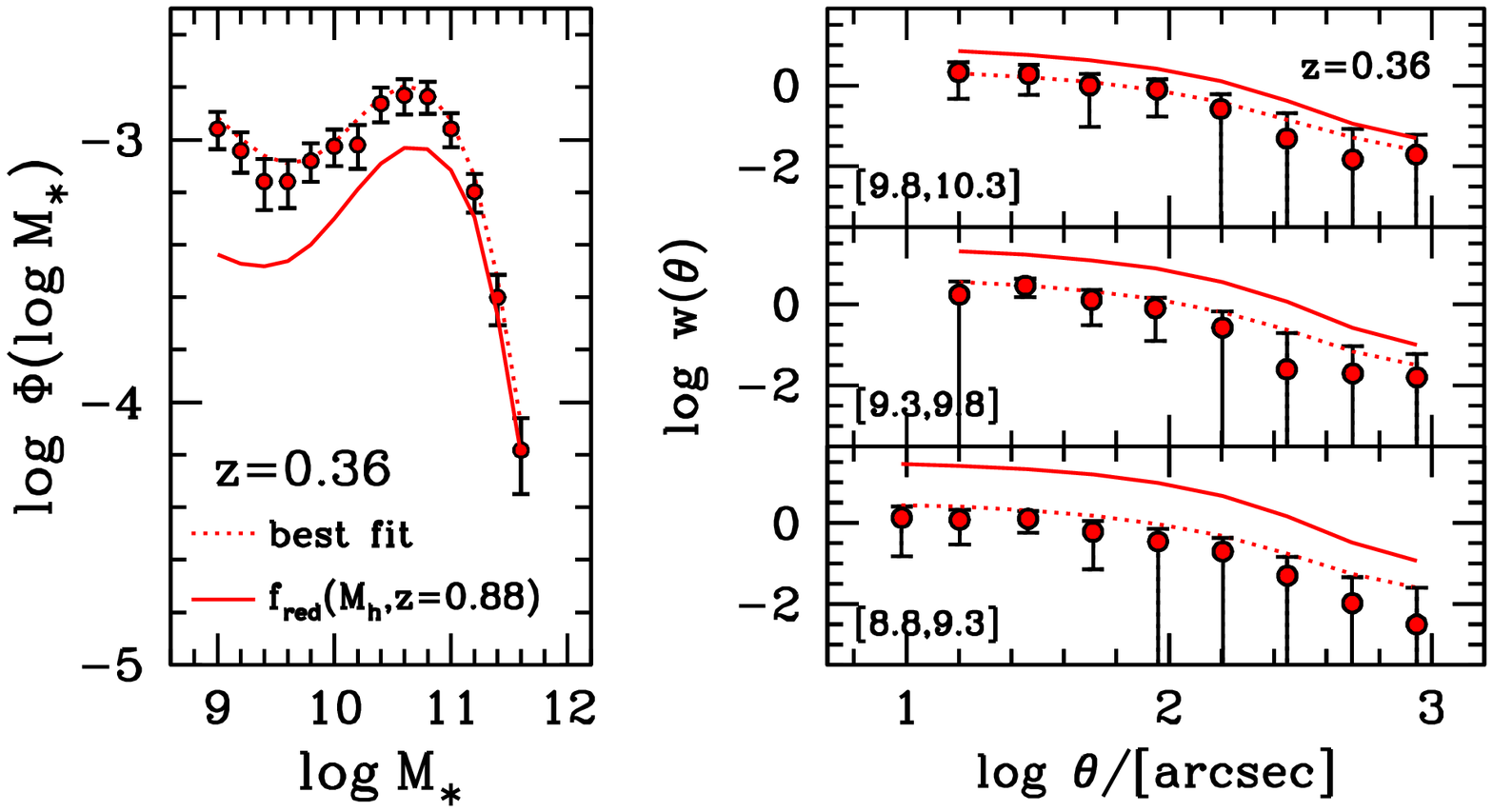}
\vspace{-7.1cm}
\caption{ \label{test_fred_evolve} Results of a model in which
  $\fred(\mhalo)$ is held fixed to the best-fit value from
  $z=0.88$. In all panels, the data are measurements from the $z=0.36$
  redshift bin. The left panel shows the SMF from $z=0.36$ along with
  the original best fit (dotted curve, taken from Figure
  \ref{bestfit_smf}). The solid curve shows the results where all
  parameters are fixed to the best-fit values except for
  $\fred(\mhalo)$, which is taken from the $z=0.88$ fit. In this
  model, $\fred(\mhalo)$ has a sharp cutoff at
  $\mhalo=10^{12}\,\msol$, thus suppressing the abundance of red
  central galaxies and lowering the overall SMF. The right panels show
  the effect on the clustering of passive galaxies. Reducing the abundance
  of quenched central galaxies increases the fraction of quenched
  galaxies that are satellites. The increased $\fsat$ enhances the
  clustering at all scales. Note that a better fit to the SMF can be
  obtained by increasing the number of red satellites, but this will
  only increase the clustering of passive galaxies. Thus, the solid curves
  should be considered a lower limit on $\wtheta$ for models in which
  $\fred(\mhalo)$ does not evolve from $z=1$ to $z=0.2$.}
\end{figure*}

Figure \ref{test_fred_evolve} demonstrates where our constraints on
the evolving population of red centrals derive from. If we assume that
the fraction of halos with red centrals is fixed at $z=0.88$, the
abundance and clustering of the overall red population is markedly
different at $z=0.36$. Figure \ref{test_fred_evolve} shows results
from the HOD model at $z=0.36$, but the five parameters of the
non-parametric $\fred(\mhalo)$ function have been replaced by the
best-fit values at $z=0.88$. In this model, the abundance of low-mass
is low by a factor of $\sim 2.5$ relative to the data, while the
clustering is too high by an order of magnitude or more. The increased
clustering amplitude is attributed to the higher satellite fraction of
passive galaxies in this model. It is possible to construct a model with
the $z=0.88$ $\fred(\mhalo)$ that relieves the tension with the SMF,
but this requires making up the difference by increasing the number of
satellite galaxies, which only increases the tension with the
clustering. In short, the only way to match both the SMF and $\wtheta$
measurements at $z=0.36$ is to increase the frequency of quenched
central galaxies relative to $z=0.88$.

%{figure}

%{figure}

%{figure}

%{figure}

\section{Discussion}

In an upcoming paper we will present a detailed analysis of the halo
occupation results presented here, comparing these results to the
growth histories, merging rates, and subhalo accretion and evolution
in high-resolution N-body simulations. But it is possible to make
significant qualitative assessment of our breakdown of the red
sequence into central and satellite galaxy components.

{\it Evolution of SHMR for high-mass galaxies.} Observations indicate
that the red sequence begins with massive galaxies at $z\gtrsim 2$
(\citealt{kriek_etal:08, williams_etal:09}). Thus it is not surprising
that massive passive and SF galaxies have substantially different
SHMRs at $z=1$. Both the halo occupation analysis presented here and
the X-ray groups in analyzed T12 indicate that, above the group halo
mass scale ($\gtrsim 10^{13} \msol$), star-forming central galaxies
are less massive than their red counterparts at fixed $\mhalo$. The
substantial difference between $\mgal$ for SF and passive subsamples
implies that star formation is not a stochastic process in these
objects: if massive central galaxies underwent periodic episodes of
star formation followed by longer-term quiescence, the galaxies at
fixed halo mass would have the same stellar mass. The results also
imply that massive quenched galaxies formed their stars very rapidly
at high redshift, essentially getting `ahead of the growth curve'
relative to central galaxies that would still be forming stars by
$z=1$. At high redshift, central galaxies essentially ``knew'' they
would be quenched by $z=1$ (see the discussion in T12).

From $z=1$ to $z=0$, the SHMRs evolve quite differently depending on
star formation activity. By $z=0.36$, the mean relations have crossed
and red central galaxies live in higher mass halos than SF central
galaxies at fixed mass. This inversion is also consistent with results
from $z=0$ studies (\citealt{mandelbaum_etal:06_gals, more_etal:11}).
Star forming galaxies of mass $10^{11} \msol$ grow by a factor of
$\sim 1.6$ using the star formation rates of \cite{noeske_etal:07a}
from $z=0.88$ to $z=0.36$. Host halos for these galaxies ($\mhalo\sim
10^{13})$ grow by a factor of $\sim 1.8$ over the same redshift
interval (\citealt{wechsler_etal:02}), thus star-forming central
galaxies grow almost as fast as their host halos. For quenched
galaxies, their growth rates are significantly slower than that of
their host halos, causing the inversion of the SHMR seen in Figure
\ref{shmr23}. Although halos will accrete substantial stellar mass
from smaller galaxies, most of this mass does not merge with the
central galaxy; this is implied by the evolution of the luminosity
function of massive passive galaxies (\citealt{wake_etal:06,
  cool_etal:08}). This mass contributed to the buildup of the
intracluster light (\citealt{conroy_etal:07, purcell_etal:07}). The
results here will put strong constraints on the growth of massive
passive galaxies in our follow-up paper.

{\it Evolution of the SHMR for low-mass galaxies.} In contrast to the
massive end of the galaxy population, low-mass galaxies show little
evolution in the SHMR as well as very little difference in the SHMR
between passive and SF subsamples. Due to the low abundance of
low-mass red central galaxies, the errors on the SHMR below the pivot
point are much higher for passive galaxies relative to SF
galaxies. For each redshift bin, the red SHMR is slightly below the SF
relation (at fixed $\mgal$), but they are consistent within the error
bars. From Figures \ref{smf_censat} and \ref{fred_evolve}, the
abundance of red centrals is nearly negligible at $z=1$ and increases
rapidly relative to other constituents of the full galaxy
population. Thus, most low mass quiescent central galaxies will be
recent additions to the red sequence and the halo masses of red
galaxies will be similar to those of SF galaxies. Low mass galaxies
have significant gas content, with $M_{\rm gas} \gtrsim \mgal$ at
$\mgal\lesssim 9.5$ at $z=0$ (\citealt{baldry_etal:08}). The
difference in stellar mass at fixed halo mass should be indicative of
the amount of this gas that has been converted into stars during the
quenching process. This proposition assumes no increase in the halo
mass; ie, the quenching mechanism is not major mergers.

\begin{figure}
\epsscale{1.2} 
\plotone{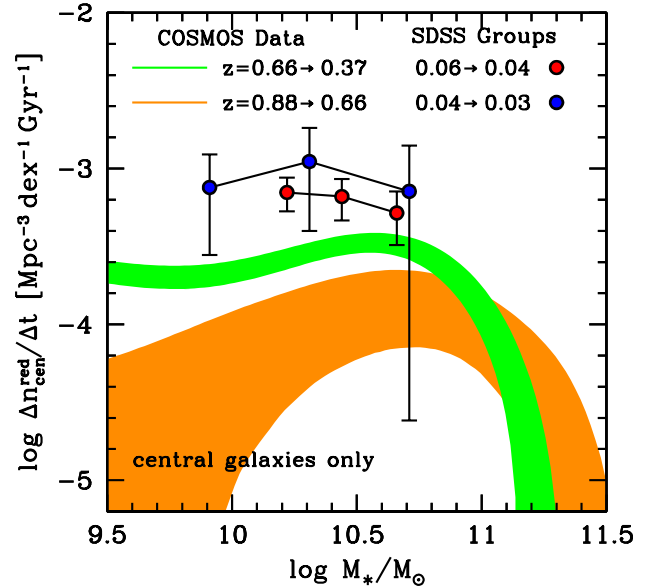}
%\vspace{-7.5cm}
\caption{ \label{redcen_rate} The rate at which central galaxies
  migrate to the red sequence as a function of stellar mass. The
  $y$-axis, $\Delta n/\Delta t$, represents the difference in the red
  central stellar mass functions between redshift bins, normalized by
  the time between each redshift. With three bins we are able to
  measure two values for $\Delta n/\Delta t$. The shaded regions show
  the 68\% confidence intervals for this quantity after combining the
  uncertainties for each redshift bin. The points with errors
  represent the same quantity but using the sample of central galaxies
  from the SDSS group catalog. The red points use a volume-limited
  sample of groups complete to $\mgal=10^{10.1}\,\msol$, with an upper
  redshift limit of 0.064. The blue points use a volume-limited sample
  of groups complete to $\mgal=10^{9.7},\msol$, with a redshift limit
  of $z=0.04$. Error bars on both sets of points are Poisson. These
  results imply that the quenching efficiency for central galaxies at
  $\mgal\lesssim 10^{10.5}\,\msol$ is increasing rapidly from $z=1$ to
  $z=0$.}
\end{figure}

  {\it The migration rate of central galaxies to the red sequence.}
  Our results are in good agreement with recent measurements from
  PRIMUS from \cite{moustakas_etal:13} that demonstrate that the
  growth of the red sequence from $z=1$ to 0 is primarily due to
  low-mass galaxies being quenched of their star formation. Our SHMR
  analysis further indicates that this growth is happening in the
  low-mass central population as opposed to satellites in groups and
  clusters. Figure \ref{redcen_rate} shows the rate at which central
  galaxies are added to the red sequence. The $x$-axis is stellar mass
  and the $y$-axis is the difference in the abundance of red centrals
  between adjacent redshift bins, divided by the time lapse between
  redshift bins (units of number/volume/dex/Gyr). The shaded regions
  shows the $1-\sigma$ range in rates given the uncertainties in the
  abundances in red centrals at each redshift. Within $1-\sigma$,
  there is evidence for an accelerated migration rate over the COSMOS
  redshift baseline, although the migration rates are consistent
  within their $2-\sigma$ uncertainties.

  Figure \ref{redcen_rate} also shows the migration rate of central
  galaxies to the red sequence using the SDSS group catalog of
  \cite{tinker_etal:11_groups}. This group catalog allows us to
  isolate only the abundance of central quenched galaxies (quenched by
  the criterion of $D_n(4000)>1.6$). Although the redshift baseline
  within the SDSS Main galaxy sample is small, the overall number of
  galaxies is very large and it is possible to detect changes in the
  abundance of red central galaxies within the volume-limited group
  catalogs of \cite{tinker_etal:11_groups} (see their Table 1). As
  discussed earlier, direct comparison of the SDSS stellar masses with
  COSMOS stellar masses is not possible, but given the lack of
  significant slope of the migration rate with stellar mass, the
  difference in $\mgal$ estimator is less relevant. The SDSS results
  yield a migration rate nearly an order of magnitude higher than the
  $z=0.88\rightarrow 0.66$ COSMOS results.

  Previous studies have also detected an acceleration of the migration
  rate onto the red sequence with cosmic time. Both the PRIMUS results
  and results from zCOSMOS of \cite{pozzetti_etal:10} find that growth
  rate, in number and mass density, for objects on the red sequence is
  increasing with decreasing redshift at $\mgal\lesssim
  10^{10.6}\msol$. These studies find significantly less evolution in
  the growth rate than in this work, which is a natural consequence of
  analyzing to overall galaxy population as opposed to focusing on
  central galaxies.

  The results here make strong predictions for the minimum $\mgal$
  that can be quenched in the field. \cite{geha_etal:12} find that
  there no isolated field galaxies below $10^9\,\msol$ that are
  passively evolving in the low-redshift NASA-Sloan Atlas. Our models
  predict that the $\fred(cen)$ drops below 1\% at
  $\mgal=3\times10^9\,\msol$ at $z=0.66$ and $\mgal=6\times
  10^9\,\msol$ at $z=0.88$. Extending this search for the minimum
  quenched field galaxy can confirm and strengthen the constraints
  from the SHMR analysis.

  {\it The quenching timescale for satellite galaxies.} The
  probability that a satellite is quenched increases monotonically
  with the time that has passed since it was accreted
  (\citealt{wetzel_etal:13_groups2}); older satellites are more likely
  to be quenched of their star formation. Thus we can compare our
  constraints on $\fred(sat)$ with our theoretical knowledge of the
  accretion and destruction of subhalos in N-body simulations. At high
  $z$, the mean age of a subhalo (i.e., the time that has elapsed
  since it was accreted) is significantly smaller than the mean age of
  subhalos at $z=0$. Dynamical friction is more efficient at higher
  redshifts because the mean density of dark matter halos increases as
  $(1+z)^3$. Making the ansatz that the oldest subhalos have the
  lowest star formation rates allows us to infer the timescale that
  must elapse for galaxies that are accreted as star-forming to
  migrate to the red sequence; e.g., if 50\% of subhalos are older
  than 4 Gyr and 50\% of satellite galaxies are red, it takes
  approximately 4 Gyr for satellite galaxies to be quenched of their
  star formation (\citealt{tinker_etal:10_drg, tinker_wetzel:10,
    wetzel_etal:13_groups2}).

\begin{figure}
\epsscale{1.2} 
\plotone{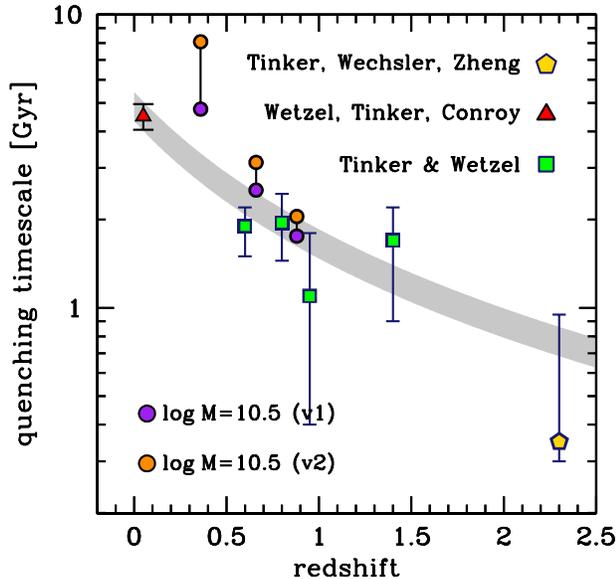}
%\vspace{-7.5cm}
\caption{ \label{quenching_time} The quenching timescale of satellite
  galaxies as a function of redshift. The purple and orange filled
  circles show results for $\mgal=10^{10.5} \msol$ galaxies in
  COSMOS. the ``v1'' method assumes that all galaxies were
  star-forming when accreted. The ``v2'' method uses $\fred(cen)$ from
  the redshift of the measurement to obtain $\tq$; these models
  bracket the physical range of models. The red triangle at $z=0.05$
  is from the analysis of SDSS groups in
  \cite{wetzel_etal:13_groups2}, which models the evolution of the red
  central fraction explicitly. The green squares are taken from the
  clustering analysis of \cite{tinker_wetzel:10}. In order of
  increasing redshift, these data points represent COMBO-17
  (\citealt{phleps_etal:06}), DEEP2 $M_B<-19.5$, DEEP2 $M_B<-20.5$
  (\citealt{coil_etal:08}), and UKIDSS-UDS
  (\citealt{williams_etal:09}). The yellow pentagon at $z=2.3$ is from
  \cite{tinker_etal:10_drg}, analyzing the clustering of DRGs from
  \cite{quadri_etal:08}. The shaded band shows $\tq \sim
  (1+z)^{-3/2}$, normalized by the datum from the SDSS groups
  data. This power-law dependence on $z$ represents the change in the
  dynamical friction timescale as the mean density of halos changes
  proportionately with the mean density of the universe. The
  observations in COSMOS as well as the other sampled plotted above
  indicate that the fraction of red satellites is constant with
  redshift. Because satellite lifetimes decrease with increasing
  redshift, the quenching of satellite galaxies must be more efficient
  in the past.}
\end{figure}

  Figure \ref{quenching_time} shows the estimated quenching time for
  $\mgal=10^{10.5} \msol$ satellite galaxies. Here we use the
  simulation results from \cite{tinker_wetzel:10}. The two values
  represent the upper and lower bounds on the quenching timescale,
  based on assumptions about the fraction of satellite galaxies that
  were quenched prior to accretion; either that $\fred(cen)=0$ or that
  $\fred(cen)$ is the value at the redshift of the measurement. In
  reality, $\fred(cen)$ will be nonzero but lower than at the redshift
  of the measurement because galaxies were accreted at higher
  redshift. We compare these results to those for $\mgal=10^{10.5}
  \msol$ galaxies $z=0$ (\citealt{wetzel_etal:13_groups2}). This
  estimate takes into account the evolution in $\fred(cen)$. We also
  show results from \cite{tinker_etal:10_drg} and
  \cite{tinker_wetzel:10} at higher redshift. These latter papers
  analyze clustering for different luminosity-defined samples, so this
  is not an apples-to-apples comparison. But in general these results
  are consistent with a scenario in which the quenching timescale of
  satellite galaxies varies with the evolving dynamical timescale of
  the host halos: $\tq\sim (1+z)^{-3/2}$.

  \cite{peng_etal:10, peng_etal:12} investigate the quenched fraction
  of galaxies as a function of local density, stellar mass, and
  redshift. They parameterize galaxy quenching as ``mass quenching''
  and ``environment quenching'', demonstrating that the effects of
  these disparate mechanisms are fully separable. ``Mass quenching''
  can be compared to central galaxy processes, while ``environment
  quenching'' is tightly associated with satellite processes. The
  fundamental difference in the approach of Peng et.~al~and this work
  is that the fundamental parameter in our approach is the mass of the
  host halo (and of the subhalo if it is a satellite), while Peng
  et.~al.~consider the stellar mass to be fundamental for the central
  galaxies and local density of galaxies to be fundamental for the
  satellites. For central galaxies, due to the small scatter between
  stellar mass and halo mass, it may not be possible to distinguish
  between these two approaches. Further work is required to see if a
  model in which central galaxy quenching is determined by galaxy mass
  fits the data as well as the model we have presented here. It is
  worth noting that \cite{peng_etal:10} find their `mass-quenching
  efficiency' to increase with cosmic time, in agreement with our
  results.

  For satellite galaxies, \cite{peng_etal:12} find that local density
  correlates better with quenching than either stellar mass of the
  satellite galaxy or host halo mass. This is at odds with our
  conclusions, as well as the model presented in
  \cite{wetzel_etal:12_groups1, wetzel_etal:13_groups2,
    wetzel_etal:13_groups3}, in which the observed correlation between
  host halo mass and quenched fraction of satellite galaxies is driven
  by the time that has elapsed since the satellites were accreted. More
  massive halos have older subhalo populations, thus contain
  satellites that are more often quenched of their star formation. In
  the next paper in this series (A. Wetzel, et.~al., in preparation),
  we model various physical mechanisms in detail, scrutinizing the
  local density model as a driver of satellite evolution.

  \cite{peng_etal:10} find no evolution with redshift in their
  environment quenching efficiency, which is in stark contrast to the
  results in Figure \ref{quenching_time} and our conclusion that
  satellite quenching efficiency is much higher in the past. The
  actual quenched fraction of satellite galaxies is nearly independent
  with redshift (cf. Figure \ref{smf_censat} and
  \cite{tinker_wetzel:10}), but \cite{peng_etal:10} do not take into
  account the redshift dependence of satellite dynamics discussed
  above, i.e., the fact that satellite at $z=1$ survive as satellites
  $\sim 1/3$ of the time $z=0$ satellites do. The \cite{peng_etal:10}
  results imply that the fraction of satellites that are quenched
  after accretion is time independent, thus their results are
  consistent with an efficiency (or timescale) that varies with the
  dynamical time of dark matter halos.

  {\it What is the mechanism responsible for the growth of the red
    sequence?} The constant $\fred(sat)$ with redshift implies that
  the rates of creation and destruction of red satellite galaxies
  roughly balance. So although the mechanisms that quench star
  formation in groups and clusters---ram pressure, strangulation,
  harassment, etc---are constantly acting on star-forming satellites
  to quench their star formation, they have minimal impact on {\it
    change} in the number of objects on the red sequence from $z=1$ to
  $z=0$\footnote{We note that the mass of these `destroyed' satellites
    is not lost, but it is likely that much of it goes into ICL and is
    not accounted for by a simple mass-weighted integral over the
    red-galaxy stellar mass function.}. The conclusion of
  \citealt{wetzel_etal:13_groups2} is that roughly 1/3 of $z=0$
  quenched galaxies with $\mgal\ge 10^{9.7}\,\msol$ were put on the
  red sequence by satellite-driven processes (their Figure 6). This is
  true whether averaging by number of galaxies or by total stellar
  mass. At $z=1$, this fraction was higher, but the overall number of
  objects on the red sequence was somewhat smaller. For central galaxies, the
  primary mechanisms proposed to quench star formation are AGN and
  major mergers, or perhaps a combination of the two as the latter may
  drive the former. To be in agreement with the results here, the
  mechanism for star formation quenching in central galaxies must
  satisfy two requirements: (1) become more efficient with time (ie,
  as $z\rightarrow 0$) and (2) be roughly independent of stellar mass.

  Let us take AGN and mergers as uncorrelated mechanisms. For mergers,
  \cite{hopkins_etal:10_mergers} find a general agreement among
  theoretical predictions and observational estimates in which the
  merger rate is $\sim 0.1$ Gyr$^{-1}$ at $z=1$ and rapidly decreases
  by a factor of $\sim 5$ from $z=1-0$ for galaxies in the range
  $10^{10}\msol<\mgal<10^{11}\msol$.  There is also a strong stellar
  mass dependence on the major merger rate (\citealt{maller:08,
    stewart_etal:09}). Which mergers actually put galaxies on the red
  sequence is not fully quantified, given that merger simulations with
  gas-rich progenitors can yield star-forming disk galaxy remnants
  (\citealt{robertson_etal:06,hopkins_etal:09}). Regarding AGN,
  although theoretical models focus on AGN as a method to halt star
  formation in massive galaxies, observed stellar mass functions of
  X-ray-AGN-hosting galaxies show little to no dependence on stellar
  mass (\citealt{bundy_etal:08, georgakakis_etal:11}). There is
  general consensus that AGN activity peaks at $z\approx 2$ and
  monotonically decreases toward $z=0$, but when quantified as a
  stellar mass function of AGN hosts, the picture is less
  clear. \cite{bundy_etal:08} show no redshift evolution in the X-ray
  AGN host SMF over $z=[0.4,1.4]$, while \cite{georgakakis_etal:11}
  find a lower amplitude of this quantity at $z\approx 0$ relative to
  $z=1$. These results rely on the pencil-thin 0.5 deg$^2$ AEGIS
  field, so sample variance may be significant. As with galaxy
  mergers, connecting AGN to quenching requires knowledge of which AGN
  matter; is there an X-ray luminosity threshold for quenching? If so,
  does it depend on stellar mass or gas mass or redshift?

  Another possibility is simply a lack of fuel for star
  formation. \cite{behroozi_etal:13_letter} demonstrate that the overall mass
  accretion rate monotonically declines for all dark matter halos at
  $z\rightarrow 0$. If baryonic accretion falls accordingly, star-forming
  central galaxies may not have a high enough surface density to
  continue forming stars

  Galaxy morphology affords an extra lever-arm in constraining power
  that we have not utilized in this paper. \cite{bundy_etal:10} find a
  population of passive disks at $z\sim 0.6$ but a paucity of such
  objects at lower $z$ (see \citealt{george_etal:13} for an
  investigation of such galaxies within groups). At low stellar
  masses, where we find the most significant increase in the red
  sequence, the morphological type with the highest fractional
  increase is ellipticals/S0, implying that the path the red sequence
  for low-mass central galaxies is accompanied by morphological change
  as well.

\section{Summary}

We have constrained the stellar to halo mass relations for passive and
star-forming galaxies over the redshift range $z=[0.2,1.0]$ in the
COSMOS field. These constraints are derived from measurements of the
stellar mass function, the angular correlation function, and
galaxy-galaxy lensing for multiple stellar mass bins within each
redshift bin. For massive galaxies, $\mgal\gtrsim 10^{10.6}\msol$, the
SHMRs for passive and SF samples exhibit significant differential
evolution, with passive galaxies growing much slower than their halos
while SF galaxies grow roughly at the same rate as their host
halos. At lower masses, there is little difference, implying that most
faint passive galaxies are recent additions to the red sequence.

Our analysis affords a breakdown of the COSMOS galaxy population into
central and satellite galaxies. With this breakdown, we demonstrate
that the number of passive satellite galaxies shows little to no
evolution with time, thus the change in the red sequence is driven by
quenching of central galaxies, primarily at low masses. The overall
migration rate of central galaxies to the red sequence is increasing
with cosmic time, with the rate at $z=0.05$ being nearly a factor of
10 higher than that derived at $z=0.78$.  Over the same redshift span,
the quenching efficiency of satellite galaxies is decreasing with
cosmic time. At $z=0.05$, the timescale for quenching is $\sim 2.5$
times longer than the quenching timescale for satellites at $z=0.88$.

We parameterize the quenching of central galaxies as being a function
of their host halo mass. At $z=0.88$, we find a sharp cutoff in
quenched central galaxies at $\mhalo\sim 10^12$, a cutoff that shifts
down 0.2-0.4 dex by $z=0.66$. These results are reminiscent of recent
theoretical work demonstrating a critical halo mass scale for
shock-heating of infalling gas; the cold-mode/hot-mode accretion
scenerio (e.g., \citealt{birnboim_dekel:03, keres_etal:05,
  keres_etal:09, dekel_birnboim:06}). This shift continues to
$z=0.36$, but at this redshift there is also a tail of quenched
central galaxies that extend to $\mhalo\lesssim 10^{11} \mhalo$, the
lowest halo mass scale for which we can probe halo occupation. The
$z=0.36$ bin does exhibit unusual clustering and abundances that
indicate sample variance is playing some role, but the redshift trends
found in the both the quenched central and satellite galaxy
populations is consistent with those found from SDSS results. Simply
removing the $z=0.36$ results from consideration does not change any
of the conclusions of this paper.

\acknowledgements We thank the referee for many helpful comments and
suggestions that have improved this work. This work was supported by
World Premier International Research Center Initiative (WPI
Initiative), MEXT, Japan. The HST COSMOS Treasury program was
supported through NASA grant HST-GO-09822. We wish to thank Tony
Roman, Denise Taylor, and David Soderblom for their assistance in
planning and scheduling of the extensive COSMOS observations.  We
gratefully acknowledge the contributions of the entire COSMOS
collaboration consisting of more than 70 scientists.  More information
on the COSMOS survey is available at {\bf
  \url{http://cosmos.astro.caltech.edu/}}. It is a pleasure the
acknowledge the excellent services provided by the NASA IPAC/IRSA
staff (Anastasia Laity, Anastasia Alexov, Bruce Berriman and John
Good) in providing online archive and server capabilities for the
COSMOS data-sets.

%%%%%%%%%%%%%%%%%%%%%%%%%%%%%%%%%%%%%%%%%%%%%%%%%%%%%%%%%%%%%%%%%%%%%%%%
%  Bibliography
%%%%%%%%%%%%%%%%%%%%%%%%%%%%%%%%%%%%%%%%%%%%%%%%%%%%%%%%%%%%%%%%%%%%%%%%

%\break
\bibliography{../risa}

\end{document}